\documentclass[prd,twocolumn,floatfix,amsmath,nofootinbib,amssymb,floatfix]{revtex4}
\usepackage{graphicx,color,dcolumn,booktabs,bm,multirow}
\usepackage{longtable,lscape}
\usepackage{txfonts}
\usepackage{overpic}
\usepackage{amssymb}
\usepackage{array}
\usepackage{indentfirst}
\usepackage{feynmf}   
\usepackage{slashed}  
\usepackage{cases}
\usepackage{color}
\usepackage{multirow}
\usepackage{epstopdf}
\usepackage{graphicx,color,dcolumn,booktabs,bm}

\graphicspath{{Figures/}} %
\usepackage[colorlinks,
                      citecolor=blue,
                      anchorcolor=red,
                      menucolor=red,
                      linkcolor=red,
                      filecolor=red,
                      runcolor=red,
                      urlcolor=blue,
                      frenchlinks=red]{hyperref}

\begin{document}

\title{Isospin violating decay of $D$ wave charmonia}
\author{Xing-Dao Guo$^{1,3}$}
\author{Qi Wu$^2$}\email{wuqi@htu.edu.cn}
\author{Dian-Yong Chen$^{4,5}$}\email{chendy@seu.edu.cn}

\affiliation{$^{1}$ College of Physics and New Energy, Xuzhou University of Technology, Xuzhou, 221111, P.R.~China}
\affiliation{$^2$ School of Physics, Henan Normal University, Xinxiang 453007, China}
\affiliation{$^3$ Department of Physics, Nankai University, Tianjin, 300071, P.R. China}
\affiliation{$^{4}$ School of Physics, Southeast University, Nanjing, 211189, P.R.~China}
\affiliation{$^5$ Lanzhou Center for Theoretical Physics, Lanzhou University, Lanzhou 730000, China}

\date{\today}

\begin{abstract}
The experimental observations of $\psi(3823)$ and $\psi(3842)$ enrich the $D$ wave charmonia family. In our previous work the decays of $\psi(3823)$ and $\psi(3842)$ to isospin conserved non-$D\bar D$ hadronic decays are well studied.
Since the isospin violated decay channel $\psi(3770)\to J/\psi \pi$ is measured clearly very recently, so in this work we fit the model parameter $\alpha$ through $BR(\psi(3770)\to J/\psi \pi)_{exp}$, then study the contribution of isospin violated decay channels of $\psi(3770)\to h_c\pi$, $\psi(3823)$ and $\psi(3842)$ to $J/\psi\pi$ and $h_c\pi$, respectively.
With this range of parameters space, the branching ratio of $\psi(3770)\to h_c\pi$ is estimated to be $(2.7\sim4.4)\times10^{-7}$.
The decay widths of $\psi(3823)$ and $\psi(3842)$ to $J/\psi\pi$ and $h_c\pi$ are also estimated.
Our results show that the partial width of $\psi(3823)\to J/\psi\pi$ is $1.12\sim1.77$ eV, the value for $\psi(3823)\to h_c\pi$ is $0.71\sim1.04$ eV, and $0.028\sim0.044$ eV for $\psi(3842)\to J/\psi\pi$, $0.34\sim0.49$ eV for $\psi(3842)\to h_c\pi$, respectively.
We hope those values could be tested by further precise measurements from the BES III, Belle II and LHCb Collaborations.
\end{abstract}

\maketitle

\section{Introduction}

Since the $\psi(3770)$ was reported by SLAC in 1977 through $e^+e^-$ annihilation~\cite{Rapidis:1977cv}, the investigation of its properties have continued.
Many works have investigated its composition~\cite{Rosner:2001nm,Voloshin:2005sd,BESIII:2023oql}, and it is widely accepted that the $\psi(3770)$ is a $^{3}D_1$ charmonium state with a very small $2^3 S_1$ component~\cite{Rosner:2004wy,Rosner:2004mi,Rosner:2001nm}. Because the $e^+e^-$ decay rate for a pure $^{3}D_1$ state is very small, the S-wave component in the $\psi(3770)$ would enhance the $e^+e^-$ decay rate and could explain its leptonic width. According to the PDG~\cite{ParticleDataGroup:2024cfk}, the mass and width of the $\psi(3770)$ are $3773.7\pm0.7$ and $27.2\pm1.0$ MeV, respectively. As the lowest-mass charmonium above the $D\bar{D}$ threshold, the dominant decay channels of $\psi(3770)$ are $D\bar D$ final states~\cite{CLEO:2005pql,CLEO:2007rrw}; this channel is OZI-allowed, leading to the width of the $\psi(3770)$ is about two orders of magnitude larger than that of charmonium below the $D\bar{D}$ threshold (such as $J/\psi$ and $\psi(3868)$).

Whether $\psi(3770)$ has appreciable non-$D\bar{D}$ decays was once a long-standing unresolved question.
Leading-order perturbative QCD (pQCD) calculations suggested that $\psi(3770)$ as relatively small non-$D\bar{D}$ branching ratios~\cite{Kuang:1989ub,Ding:1991vu,Rosner:2001nm,Rosner:2004wy,Eichten:2007qx,Voloshin:2005sd}.
The first non-$D\bar{D}$ decay channel of the $\psi(3770)$ is $J/\psi \pi^+ \pi^-$, which was observed by the BES Collaboration in 2003~\cite{BES:2003bes} and confirmed by the CLEO Collaboration~\cite{CLEO:2005zky} as well as other hadronic transitions, including $J/\psi \pi^0 \pi^0$, $\eta J/\psi$~\cite{CLEO:2005zky}, the radiative transitions $\gamma \chi_{cJ}(J=0,1)$~\cite{CLEO:2005vma,CLEO:2006nor}, and the decay into light hadrons $\phi\eta$~\cite{CLEO:2005zrs}.
Further measurements by BES II and BES III showed that decays into final states other than $D\bar{D}$ make up a substantial fraction, $(15.1\pm5.6\pm1.8)\%$, of the total hadronic width~\cite{BES:2006fpf,BES:2006dso,BES:2007cev,BES:2008vad}.

The non-$D\bar{D}$ decays of the $\psi(3770)$ have sparked extensive and intense discussions~\cite{Kuang:1989ub,Ding:1991vu,Rosner:2001nm,Rosner:2004wy,Eichten:2007qx,Voloshin:2005sd,Achasov:2005qb,He:2008xb,CLEO:2005pql,CLEO:2005mpm,Achasov:1990gt,Achasov:1994vh}.
Among these decays, the hadronic transition $\psi(3770)\to \pi^0 J/\psi$ is one of the important channels for understanding non-$D\bar{D}$ decays and testing model predictions for the $\psi(3770)$~\cite{Voloshin:2005sd,Ioffe:1980mx,Zhang:2009kr,Guo:2012tj,Achasov:2005qb,Liu:2009dr}.
As suggested in Ref.~\cite{Voloshin:2005sd}, it is crucial to establish whether the decay $\psi(3770)\to \pi^0 J/\psi$ conserves or violates isospin, because a non-negligible four-quark component with isospin $I=1$ in the $\psi(3770)$ could significantly enhance this decay.
The observation of unexpectedly large non-$D\bar{D}$ decay rates for the $\psi(3770)$ implies that additional mechanisms contribute.
One plausible non-perturbative candidate is the hadronic loop mechanism, which may be responsible for these anomalously large non-$D\bar{D}$ decay modes~\cite{Zhang:2009kr,Guo:2012tj,Achasov:2005qb,Liu:2009dr,Wang:2011yh,Li:2013zcr} (for a comprehensive review of the non-perturbative aspects of charmoniam, we refer the reader to a recent review~\cite{Bai:2026atm}).

Based on the fact that $\psi(3770)$ has a mass just above the open $D\bar{D}$ threshold, the authors of Ref.~\cite{Zhang:2009kr} proposed that the dominant $D\bar{D}$ decay is strongly correlated with the non-$D\bar{D}$ decays.
By taken into account final state interaction (FSI), Ref.~\cite{Liu:2009dr} calculated the branching ratios of $\psi(3770)\to D\bar{D}\to J/\psi\eta$, $\rho\pi$, $\omega\eta$, and $K^*K$, and the results indicates that FSI is obviously non-negligible.
In Ref.~\cite{Guo:2012tj}, the authors estimate the decay $\psi(3770)\to J/\psi\pi$.
In their article, they calculate the decay channel $\psi(3770)\to J/\psi \pi$ through an onshell $D$ meson loop and consider the contributions from both the effective coupling constants between the neutral and charged decay channels and the mass difference between $D^0$ and $D^\pm$.
During their study, since the $\psi(3770)$ is just above the $D\bar D$ threshold, they only calculate the contribution of the dispersive part of the $D\bar D(D^*)$ loop to $\psi(3770)\to J/\psi \pi$.


Very recently, the BES III Collaboration measured the branching ratio of $\psi(3770)\to J/\psi \pi$ precisely and gave its value as $(2.08\pm0.36\pm0.25\pm0.21)\times10^{-5}$ with different uncertainties~\cite{BESIII:2026pyf}.
The experimentally measured value is in good agreement with the calculations based on the hadronic loop mechanism, highlighting the importance of hadronic loop effects in this isospin-violating decay.
Moreover, in 2013 the Belle Collaboration found a new resonance $\psi(3823)$ with a mass of $3823.1$ MeV~\cite{Belle:2013ewt}.
In 2019, the LHCb Collaboration observed a charmonium state $\psi(3842)$ with a mass around $3842.7$ MeV~\cite{LHCb:2019lnr}.
The observation of these two states promote us to consider $\psi(3823)$ and $\psi(3842)$ as higher excited states of $\psi(3770)$, which are regarded as members of the D-wave triplet charmonium states with $J=1$, $2$, and $3$ in the present work, respectively.
In our following study, we denote them as $\psi_1$, $\psi_2$, and $\psi_3$ for simplicity.
In our pervious work~\cite{Qi:2025hwd,Qi:2025xkz}, we have estimated the non-$D\bar D$ hadronic decays of D-wave charmonia, focusing on isospin-conserved decay channels.
We note that there are many isospin-violating decay channels in the D-wave charmonium family, and some of them have been observed in experiment.

As is known, the isospin-breaking effect comes from the mass difference between up and down quarks.
Under the up-down quark symmetry limit, the neutral and charged currents will cancel each other, but the tiny difference between them will lead to the violation.
There are three concrete ways.
One is through the tiny difference in the effective coupling constants between neutral and charged decay channels.
Another is through the mass difference between charged and neutral particles (such as $m_{D^0}\neq m_{D^\pm}$).
The third is through light meson mixing such as $\pi-\eta$ mixing.

In this work, we employ the effective Lagrangian approach to investigate the isospin-violating decays of D-wave charmonia with the hadronic loop mechanism.
Different from Ref.~\cite{Liu:2009dr}, we not only consider the vertex $\psi_1 D\bar D$, which relate the D wave charmonia to the intermediate charmed mesons in the loop, but also include the vertexes $\psi_1 D^*\bar D$ and $\psi_1 D^*\bar D^*$ that may contribute to the isospin-violating decay processes.
In addition, we calculate the isospin-violating processes under three scenarios, namely, the effective coupling constants $g_{\psi_1 D^0\bar D^0}\neq g_{\psi_1 D^\pm\bar D^\mp}$, the mass difference $m_{D^0}\neq m_{D^\pm}$, and the $\pi-\eta$ mixing, respectively.
As for the effects of $g_{D^{*0} D^0\pi}\neq g_{D^{*\pm} D^\pm\pi}$, although the exactly experimental data $BR(D^{*0} \to D^0 +\pi)_{exp}$,
$BR(D^{*\pm} \to D^\pm +\pi)_{exp}$, and $BR(D^{*\pm}_s \to D^\pm_s +\pi)_{exp}$ are measured, the total decay widths of $D^{*0}$ and  $D^{*}_s$ are not measured accurately~\cite{ParticleDataGroup:2024cfk}.
Thus, in this study we neglect the difference between them and take them to be the same value under the heavy-quark limit and chiral symmetry.

In our pervious study, we calculated the decay widths of $\psi(3823)$ and $\psi(3842)$ to light vector and pseudoscalar mesons final states in Ref.~\cite{Qi:2025xkz},
and calculated the decay widths of $\psi(3823)$ and $\psi(3842)$ to $J/\psi\eta$ and $\eta_c \omega$ in Ref.~\cite{Qi:2025hwd}.
Benefiting from the experimental measurement of the branching fraction $\psi(3770)\to \pi^0 J/\psi$ by the BES III Collaboration, the phenomenological parameter in our model can be fixed, and we will further study the decays of $\psi(3823)$ and $\psi(3842)$ into $J/\psi\pi$ and $h_c \pi$.
Based on the experimental data, it is noted that $\psi(3770)$ and $\psi(3842)$ can couple to $D\bar D$ through P and F-waves, respectively,
while the coupling between $\psi(3823)$ and $D\bar D$ is forbidden due to the conservations of $J^P$ quantum numbers.
Thus, the charmed meson loop will be the dominant contribution to the hidden charm decay processes of $\psi(3823)$ and $\psi(3842)$, and in our present study, we perform a systematic investigation of the isospin-violating two-body hidden charm decay processes of the D-wave charmonia $\psi(3770)$, $\psi(3823)$ and $\psi(3842)$, and try to provide further insight into the isospin-violating decay properties of the D-wave charmonia.

This work is organized as follows. After the introduction, we list the corresponding effective Lagrangians and then calculate the Feynman amplitudes and branching ratios of the decay channels of D-wave charmonia through meson loop mechanism in section II.
Then, we fit the parameter through the experimental data and estimate the corresponding numerical results in section III.
The last section is devoted to a short summary.

\section{Theoretical framework }

\subsection{ Effective Lagrangians}
As mentioned in the previous section, charmed meson loops play an important rule in understanding the non-$D\bar{D}$ decays of the $\psi(3770)$, especially the isospin-violating decay channels.
In the heavy quark limit, the wave functions of the charmed mesons are independent of the spin and flavor of the heavy quark,
and the S-wave charmed mesons can be expressed in matrix form~\cite{Mannel:1997ky,Brambilla:2004jw,Colangelo:2003sa}
\begin{eqnarray}
H_1&=&\frac{1+v\!\!\!\slash}{2}\left[\mathcal{D}^{\ast\mu}\gamma_\mu-\mathcal{D}\gamma_5\right], \nonumber
\\
H_2&=&\left[\bar{\mathcal{D}}^{\ast\mu}\gamma_\mu-\bar{\mathcal{D}}\gamma_5\right]\frac{1-v\!\!\!\slash}{2},
\\
\bar{H}_{1,2} &=&\gamma^0 H_{1,2}^\dagger \gamma^0 \nonumber
\end{eqnarray}
with $\mathcal{D}^{(\ast)}=(D^{(\ast)0}, D^{(\ast) +}, D_s^{(\ast)+})$ for simplicity.

For heavy quarkonium, the S-wave, P-wave, and D-wave charmonia can be expressed as~\cite{Mannel:1997ky,Brambilla:2004jw,Colangelo:2003sa,Lipkin:1986av,Casalbuoni:1996pg,Kaymakcalan:1983qq,Oh:2000qr}
\begin{eqnarray}
\mathcal{R}&=&\frac{1+v\!\!\!\slash}{2} \left[\psi^{\mu}\gamma_\mu-\eta_{c}\gamma_5 \right]\frac{1-v\!\!\!\slash}{2}, \nonumber\\
\mathcal{P}^{\mu}&=&\frac{1+v\!\!\!\slash}{2}\left[ \chi^{\mu\alpha}_2 \gamma_\alpha+
\frac{1}{\sqrt{2}}\varepsilon^{\mu\alpha\beta\gamma} v_\alpha \gamma_\beta \chi_{2\gamma}\right. \nonumber\\
		&&+\left.\frac{1}{\sqrt{3}}(\gamma^\mu-v^\mu)\chi_0+h_c^\mu\gamma_5 \right] \frac{1-v\!\!\!\slash}{2},
\\
\mathcal{J}^{\mu\lambda}&=&\frac{1+v\!\!\!\slash}{2}\left[\psi^{\mu\alpha\lambda}_3 \gamma_\alpha+\frac{1}{\sqrt{6}}\Big(\varepsilon^{\mu\alpha\beta\rho}v_\alpha \gamma_\beta \psi^\lambda_{2\rho}+\varepsilon^{\lambda\alpha\beta\rho}v_\alpha \gamma_\beta \psi^\mu_{2\rho}\Big)\right. \nonumber\\
		&&+\left.\frac{\sqrt{15}}{10}\Big((\gamma^\mu-v^\mu)\psi^\lambda_1+(\gamma^\lambda-v^\lambda)\psi^\mu_1\Big)\right.\nonumber\\
		&&-\left.\frac{1}{\sqrt{15}}\Big(g^{\mu\lambda}-v^\mu v^\lambda\Big)\gamma_\alpha \psi^\alpha_1+\eta^{\mu\lambda}_{c2}\gamma_5 \right]
\frac{1-v\!\!\!\slash}{2}.\nonumber
\end{eqnarray}
Then the general Lagrangians describing the couplings of the S-wave, P-wave, and D-wave charmonia to a pair of charmed and anti-charmed mesons can be written as~\cite{Mannel:1997ky,Brambilla:2004jw,Colangelo:2003sa,Lipkin:1986av,Casalbuoni:1996pg,Kaymakcalan:1983qq,Oh:2000qr}
\begin{eqnarray}
\mathcal{L}_{\mathcal{S}}&=&ig_sTr[\mathcal{R} \bar H_2\stackrel{\leftrightarrow}{\partial_{\mu}}\gamma_\mu \bar H_1]+\rm{h.c.}, \nonumber
\\
\mathcal{L}_{\mathcal{P}}&=&ig_pTr[\mathcal{P}^\mu \bar H_2\gamma_\mu \bar H_1]+\rm{h.c.},
\\
\mathcal{L}_{\mathcal{D}}&=&g_dTr[\mathcal{J}^{\mu\nu}\bar H_2 \stackrel{\leftrightarrow}{\partial_{\mu}}\gamma_\lambda \bar H_1]+\rm{h.c.}. \nonumber
\label{Lagrangian}
\end{eqnarray}
After further expanding the Lagrangian mentioned above, we can derive the specific form of the Lagrangians for the coupling vertices involving charmonia and charmed mesons under the above Lagrangian framework.

The effective Lagrangians for the coupling vertices involving charmonia and charmed mesons have been widely studied and used in many works~\cite{Mannel:1997ky,Brambilla:2004jw,Colangelo:2003sa}.
Here we just list the corresponding Lagrangians describing the interactions between S-, P-, and D-wave charmonia and charmed mesons as follows
\begin{eqnarray}
\mathcal{L}_{\psi \mathcal{D}^{(*)}\mathcal{D}^{(*)}}&=&
-ig_{\psi \mathcal{D} \mathcal{D}}
\psi_{\mu}\Big(\partial^{\mu}\mathcal{\mathcal{D}\mathcal{D}}^{\dagger}-\mathcal{D}\partial^{\mu}\mathcal{D}^{\dagger}\Big)\nonumber \\
		&&+g_{\psi \mathcal{D}^* \mathcal{D}}\varepsilon^{\mu\nu\alpha\beta}\partial_{\mu}\psi_{\nu}
\Big(\mathcal{D}_{\alpha}^{*}\stackrel{\leftrightarrow}{\partial}_{\beta}\mathcal{D}^{\dagger}
-\mathcal{D}\stackrel{\leftrightarrow}{\partial_{\beta}} \mathcal{D}_{\alpha}^{*\dagger}\Big) \nonumber\\
		&&+ig_{\psi \mathcal{D}^* \mathcal{D}^*}
\psi^\mu\Big(\mathcal{D}_\nu^*\stackrel{\leftrightarrow}{\partial^\nu} \mathcal{D}_\mu^{*\dagger}+\mathcal{D}_\mu\stackrel{\leftrightarrow}{\partial^\nu} \mathcal{D}_\nu^{*\dagger}\nonumber\\
		&&-\mathcal{D}_\nu \stackrel{\leftrightarrow}{\partial_\mu} \mathcal{D}^{*\dagger\nu}\Big),\nonumber\\
\mathcal{L}_{h_c \mathcal{D}^{(*)}\mathcal{D}^{(*)}}&=&
g_{h_c \mathcal{D}^*\mathcal{D}}
h_c^{\mu}\Big(\mathcal{\mathcal{D}\mathcal{D}}^{*\dagger}_\mu+\mathcal{D}^*_{\mu}\mathcal{D}^{\dagger}\Big)\nonumber \\
		&&+ig_{h_c \mathcal{D}^* \mathcal{D}^*}
\varepsilon^{\mu\nu\alpha\beta}\partial_{\mu}h_{c\nu} \mathcal{D}_{\alpha}^{*}\mathcal{D}^{*\dagger}_{\beta}, \\
\mathcal{L}_{\psi_J \mathcal{D}^{(*)}\mathcal{D}^{(*)}}&=&
g_{\psi_1 \mathcal{D} \mathcal{D}}
\psi^\mu_1 \Big(\mathcal{D} \partial_\mu \mathcal{D}^\dagger-\mathcal{D}^\dagger\partial_\mu\mathcal{D}\Big)\nonumber\\
		&&+g_{\psi_1 \mathcal{D}^* \mathcal{D}}
\varepsilon^{\mu\nu\alpha\beta}\Big(\mathcal{D}{\stackrel{\leftrightarrow}{\partial_\mu}}\mathcal{D}^{\ast\dagger}_\beta
-\mathcal{D}^{\ast}_\beta{\stackrel{\leftrightarrow}{\partial_\mu}}\mathcal{D}^\dagger\Big)\partial_\nu \psi_{1\alpha}\nonumber\\
		&&+g_{\psi_1 \mathcal{D}^\ast \mathcal{D}^\ast}
\left[-4\Big(\psi^\mu_1 \mathcal{D}^{\ast\nu\dagger}\partial_\mu \mathcal{D}^{\ast}_\nu
-\psi^\mu_1 \mathcal{D}^{\ast}_\nu \partial_\mu \mathcal{D}^{\ast\nu\dagger}\Big)\right.\nonumber\\
		&&+\left.\psi^\mu_1 \mathcal{D}^{\ast\nu\dagger}\partial_\nu \mathcal{D}^{\ast}_\mu
-\psi^\mu_1 \mathcal{D}^{\ast\nu}\partial_\nu \mathcal{D}^{\ast\dagger}_\mu\right]\nonumber\\
		&&+ig_{\psi_2 \mathcal{D}^* \mathcal{D}}
\psi^{\mu\nu}_2\Big(\mathcal{D}{\stackrel{\leftrightarrow}{\partial_\nu}}\mathcal{D}^{\ast\dagger}_\mu
-\mathcal{D}^{\ast}_\mu{\stackrel{\leftrightarrow}{\partial_\nu}}\mathcal{D}^\dagger\Big)\nonumber\\
		&&+ig_{\psi_2 \mathcal{D}^\ast \mathcal{D}^\ast}
\varepsilon_{\alpha\beta\mu\nu}\Big(\mathcal{D}^{\ast\nu}{\stackrel{\leftrightarrow}{\partial^\beta}}\mathcal{D}^{\ast\dagger}_\lambda
-\mathcal{D}^{\ast\nu\dagger}{\stackrel{\leftrightarrow}{\partial^\beta}}\mathcal{D}^{\ast}_\lambda\Big)
\partial^\mu \psi^{\alpha\lambda}_2\nonumber\\
		&&+g_{\psi_3 \mathcal{D}^\ast \mathcal{D}^\ast}
\psi^{\mu\nu\alpha}_3 \Big(\mathcal{D}^{\ast}_\alpha{\stackrel{\leftrightarrow}{\partial_\mu}}\mathcal{D}^{\ast\dagger}_\nu
+\mathcal{D}^{\ast}_\nu{\stackrel{\leftrightarrow}{\partial_\mu}}\mathcal{D}^{\ast\dagger}_\alpha\Big).\nonumber
	\end{eqnarray}

Under the heavy quark limit and the chiral symmetry, we can write the Lagrangian describing the interactions between light meson and charmed mesons. Thus we have~\cite{Falk:1992cx,Chen:2014sra,Yan:1992gz,Cheng:1992xi,Wise:1992hn}
\begin{eqnarray}
\mathcal{L}_{\mathcal{D}^* \mathcal{D}^{(*)}\mathcal{P}}&=&-ig_{\mathcal{D}^\ast \mathcal{D}
\mathcal{P}}\Big(\mathcal{D}^{i\dagger} \partial^\mu \mathcal{P}_{ij}\mathcal{D}^{\ast j}_\mu
-\mathcal{D}^{\ast i\dagger}_\mu \partial^\mu \mathcal{P}_{ij}\mathcal{D}^{j}\Big)\nonumber\\
		&&+\frac{1}{2}g_{\mathcal{D}^\ast \mathcal{D}^\ast \mathcal{P}}\varepsilon_{\mu\nu\alpha\beta}\mathcal{D}^{\ast \mu\dagger}_i \partial^\nu \mathcal{P}_{ij}{\stackrel{\leftrightarrow}{\partial^\alpha}}\mathcal{D}^{\ast \beta}_j,		
\end{eqnarray}
with $\mathcal{P}$ being the pseudoscalar meson and its matrix form is~\cite{Casalbuoni:1996pg,Casalbuoni:1992gi,Casalbuoni:1992dx}
\begin{eqnarray}
 \mathcal{P}&=&\left(\begin{array}{ccc}
\frac{\pi^{0}}{\sqrt{2}}+\alpha' \eta+\beta \eta \prime & \pi^{+} & K^{+} \\
\pi^{-} & -\frac{\pi^{0}}{\sqrt{2}}+\alpha' \eta+\beta \eta \prime & K^{0} \\
K^{-} & \bar{K}^{0} & \gamma \eta+\delta \eta^{\prime}
		\end{array}\right), \nonumber
\end{eqnarray}
where the parameters $\alpha'$, $\beta$, $\gamma$, and~$\delta$ are parameters that represent the mixing between $\eta$, $\eta'$, $\eta_0$, and $\eta_8$.
The four parameters can be related to the mixing angle $\theta$ and are defined as~\cite{Gilman:1987ax},
\begin{eqnarray}
\alpha'=\frac{\mathrm{cos}\theta-\sqrt{2}\mathrm{sin}\theta}{\sqrt{6}}&, \quad& \beta=\frac{\mathrm{sin}\theta+\sqrt{2}\mathrm{cos}\theta}{\sqrt{6}}, \nonumber\\
\gamma=\frac{-2\mathrm{cos}\theta-\sqrt{2}\mathrm{sin}\theta}{\sqrt{6}}&, \quad& \delta=\frac{-2\mathrm{sin}\theta+\sqrt{2}\mathrm{cos}\theta}{\sqrt{6}}, \quad
\end{eqnarray}
where $\theta$ is the mixing angle. In the present work, we adopt the mixing angle $\theta=19.1^{\circ}$~\cite{MARK-III:1988crp, DM2:1988bfq}.

The effective Lagrangian for $\eta-\pi$ mixing reads~\cite{Xiao:2016hoa,Yue:2025wcl}
\begin{eqnarray}
\mathcal{L}_{\eta\pi^0}=\mu \dfrac{m_d-m_u}{\sqrt 3}\eta\pi^0,
\end{eqnarray}
where $m_d$ and $m_u$ are the current quark masses of the d and u quarks, $m_\pi^2=(m_d + m_u)\mu$, $m_\eta^2=\frac{2}{3}(m + m_s)\mu$, $m=(m_d + m_u)/2$, and $\mu$ is the condensate parameter.

\subsection{ Amplitudes for $\psi(3770)\to J/\psi\pi$}

\begin{figure}[http]
\includegraphics[width=8.2cm]{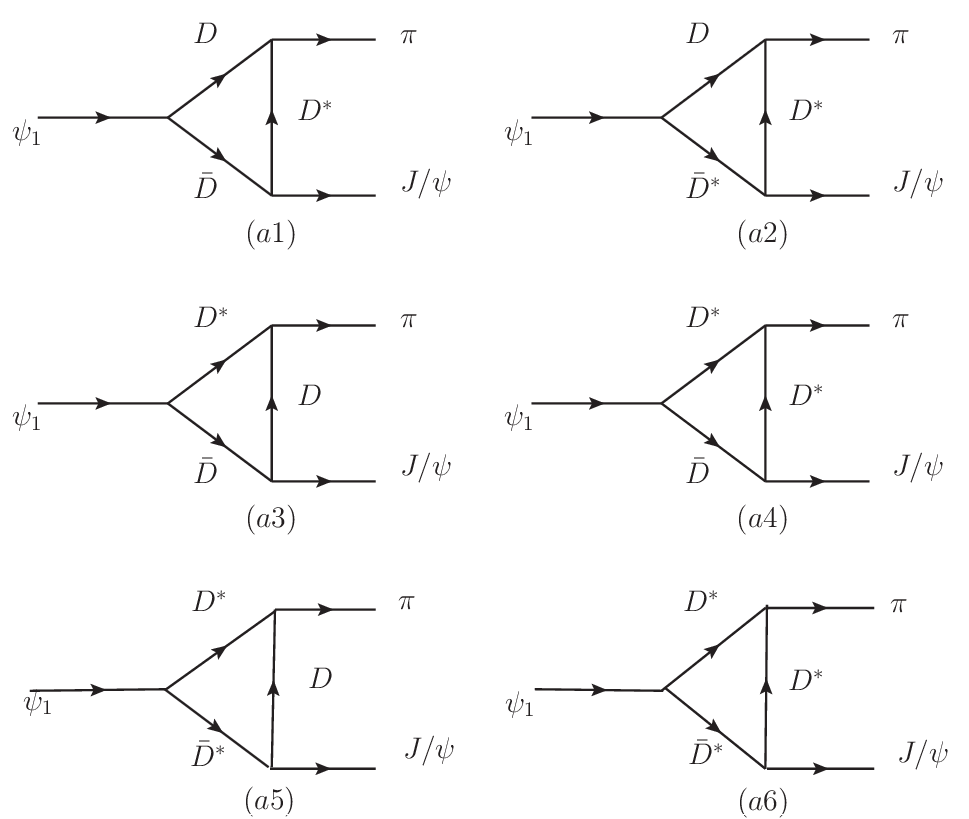}
\caption{Feynman diagrams contributing to $\psi(3770)\to J/\psi\pi$ at the hadronic level.\label{fig1}}
\end{figure}

Utilizing the effective Lagrangians mentioned above, we can obtain the hidden charm decay process of the D-wave charmonium $\psi(3770)$ to $J/\psi\pi$.
In Fig.~\ref{fig1} we list all the Feynman diagrams for $\psi(3770)$ decaying directly to $J/\psi\pi$.

\begin{figure}[http]
\includegraphics[width=8.2cm]{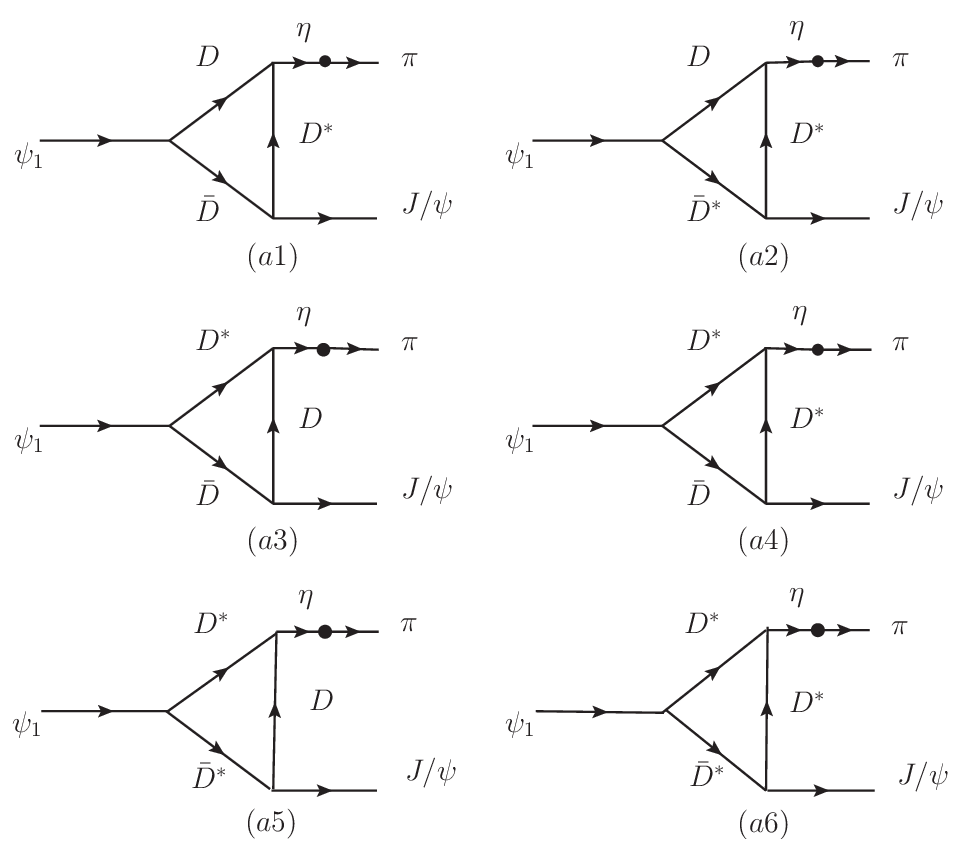}
\caption{Feynman diagrams contributing to $\psi(3770)\to J/\psi\pi$ via $\eta$-$\pi$ mixing at the hadronic level.\label{fig1mix}}
\end{figure}

As mentioned above, the $\eta$-$\pi$ mixing can also contribute to $\psi(3770)\to J/\psi\pi$. Thus, the corresponding diagrams are also listed in
Fig.~\ref{fig1mix}.

We obtain the amplitudes corresponding to Fig.~\ref{fig1} as
\begin{equation}
\begin{array}{rl}
\mathcal{M}_{a1}= &i^3\displaystyle\int \dfrac{d^4 q}{(2\pi)^4}[-i g_{\psi_1 DD}(-i)(-1)(p_1-p_2)\cdot\epsilon_{\psi_1}(p)]\\
& [-i g_{D^*D\pi}(-ip_3)^\sigma] [g_{\psi D^*D} \epsilon_{\mu\nu\alpha\beta}(-i p_4)^\mu \epsilon_\psi^\nu(p_4)\\
&(-i)(q-p_2)^\alpha]
\dfrac{1}{p_1^2-m_D^2}\dfrac{1}{p_2^2-m_D^2}\\
&\dfrac{-g^{\sigma\beta}+q^\sigma q^\beta/m_{D^*}^2}{q^2-m_{D^*}^2},
\end{array}
\end{equation}

\begin{equation}
\begin{array}{rl}
\mathcal{M}_{a2}= &i^3\displaystyle\int \dfrac{d^4 q}{(2\pi)^4}
[g_{\psi_1 D^*D} \epsilon_{\mu\nu\alpha\sigma} (-ip)^\mu  \epsilon_{\psi_1}^\nu(p) (-i)(p_1-p_2)^\alpha]\\
& [-i g_{D^*D\pi}(-i p_3)^\tau] (i g_{\psi D^* D^*} )\\
&[(-1)(-i)(-\epsilon_\psi(p_4)\cdot (q-p_2)g^{\beta\chi} \\
&+\epsilon_\psi^\beta(p_4)(q-p_2)^\chi +\epsilon_\psi^\chi(p_4)(q-p_2)^\beta)] \\
&\dfrac{1}{p_1^2-m_D^2}\dfrac{-g^{\sigma\beta}+p_2^\sigma p_2^\beta/m_{D^*}^2}{p_2^2-m_{D^*}^2}
\dfrac{-g^{\tau\chi}+q^\tau q^\chi/m_{D^*}^2}{q^2-m_{D^*}^2},
\end{array}
\end{equation}

\begin{equation}
\begin{array}{rl}
\mathcal{M}_{a3}= &i^3\displaystyle\int \dfrac{d^4 q}{(2\pi)^4}
[g_{\psi_1 D^*D} \epsilon_{\mu\nu\alpha\beta}(-1)\\
& (-ip)^\mu  \epsilon_{\psi_1}^\nu(p) (-i)(p_1-p_2)^\alpha]\\
& [-i g_{D^*D\pi}(-1)(-i p_3)^\sigma]
[-i g_{\psi DD}(-i)(q-p_2)\cdot\epsilon_\psi(p_4)]\\
&\dfrac{-g^{\sigma\beta}+p_1^\sigma p_1^\beta/m_{D^*}^2}{p_1^2-m_{D^*}^2}\dfrac{1}{p_2^2-m_D^2}\dfrac{1}{q^2-m_D^2},
\end{array}
\end{equation}

\begin{equation}
\begin{array}{rl}
\mathcal{M}_{a4}= &i^3\displaystyle\int \dfrac{d^4 q}{(2\pi)^4} \\
&[g_{\psi_1 D^*D} \epsilon_{\mu\nu\alpha\theta}(-1) (-ip)^\mu  \epsilon_{\psi_1}^\nu(p) (-i)(p_1-p_2)^\alpha]\\
&[\dfrac{g_{D^*D^*\pi} }{2}\epsilon_{\chi\lambda\sigma\rho}(-1)(-i p_3)^\lambda(-i)(p_1+q)^\sigma g^{\rho\beta}] \\
& [g_{\psi D^*D} \epsilon_{\kappa\tau\phi\xi} (-i p_4)^\kappa \epsilon_\psi^\tau(p_4)(-i)(q-p_2)^\phi]\\
&\dfrac{-g^{\theta\beta}+p_1^\theta p_1^\beta/m_{D^*}^2}{p_1^2-m_{D^*}^2}\dfrac{1}{p_2^2-m_D^2}
\dfrac{-g^{\xi\chi}+q^\xi q^\chi/m_{D^*}^2}{q^2-m_{D^*}^2},
\end{array}
\end{equation}

\begin{equation}
\begin{array}{rl}
\mathcal{M}_{a5}= &i^3\displaystyle\int \dfrac{d^4 q}{(2\pi)^4}
[i g_{\psi_1 D^*D^*}(-i)(-4\epsilon_{\psi_1}(p)\cdot (p_1-p_2)g^{\alpha\beta} \\
&+\epsilon_{\psi_1}^\beta(p)(p_1-p_2)^\alpha +\epsilon_{\psi_1}^\alpha(p)(p_1-p_2)^\beta)]\\
& [-i g_{D^*D\pi}(-1)(-i p_3)^\sigma] \\
& [g_{\psi D^*D} \epsilon_{\kappa\tau\phi\xi} (-i p_4)^\kappa \epsilon_\psi^\tau(p_4) (-i)(-1)(q-p_2)^\phi ]\\
&\dfrac{-g^{\sigma\beta}+p_1^\sigma p_1^\beta/m_{D^*}^2}{p_1^2-m_{D^*}^2}
\dfrac{-g^{\xi\alpha}+p_2^\xi p_2^\alpha/m_{D^*}^2}{p_2^2-m_{D^*}^2}\dfrac{1}{q^2-m_D^2},
\end{array}
\end{equation}

\begin{equation}
\begin{array}{rl}
\mathcal{M}_{a6}= &i^3\displaystyle\int \dfrac{d^4 q}{(2\pi)^4}  [i g_{\psi_1 D^*D^*}(-i)
(-4\epsilon_{\psi_1}(p)\cdot (p_1-p_2)g^{\alpha\beta}\\
&+\epsilon_{\psi_1}^\beta(p)(p_1-p_2)^\alpha +\epsilon_{\psi_1}^\alpha(p)(p_1-p_2)^\beta)]\\
& (\dfrac{g_{D^*D^*\pi} }{2} ) [\epsilon_{\tau\lambda\sigma\rho}(-1)(-i p_3)^\lambda(-i)(p_1+q)^\sigma ] \\
& [i g_{\psi D^*D^*}(-i)(-1)(-\epsilon_{\psi}(p_4)\cdot (q-p_2)g^{\theta\chi}\\
&+\epsilon_\psi^\chi(p_4)(q-p_2)^\theta +\epsilon_\psi^\theta(p_4)(q-p_2)^\chi)]\\
&\dfrac{-g^{\rho\beta}+p_1^\rho p_1^\beta/m_{D^*}^2}{p_1^2-m_{D^*}^2}
\dfrac{-g^{\theta\alpha}+p_2^\alpha p_2^\theta/m_{D^*}^2}{p_2^2-m_{D^*}^2} \\
&\dfrac{-g^{\tau\chi}+q^\tau q^\chi/m_{D^*}^2}{q^2-m_{D^*}^2}.
\end{array}
\end{equation}

The amplitudes contributing to the $\psi(3770)\to J/\psi\pi$ process through $\eta$-$\pi$ mixing, as shown in Fig.~\ref{fig1mix}, can be obtained by employing the following substitutions
\begin{equation}
\begin{array}{rl}
\mathcal{M}_{a1}^{mix}&= \left.\mathcal{M}_{a1}\dfrac{\sqrt 3}{4}\dfrac{m_d-m_u}{m_s-m}\right|_{g_{DD^*\pi} \to g_{DD^*\eta}},\nonumber\\
\mathcal{M}_{a2}^{mix}&= \left.\mathcal{M}_{a2}\dfrac{\sqrt 3}{4}\dfrac{m_d-m_u}{m_s-m}\right|_{g_{DD^*\pi} \to g_{DD^*\eta}},\nonumber\\
\mathcal{M}_{a3}^{mix}&= \left.\mathcal{M}_{a3}\dfrac{\sqrt 3}{4}\dfrac{m_d-m_u}{m_s-m}\right|_{g_{DD^*\pi} \to g_{DD^*\eta}},\nonumber\\
\mathcal{M}_{a4}^{mix}&= \left.\mathcal{M}_{a4}\dfrac{\sqrt 3}{4}\dfrac{m_d-m_u}{m_s-m}\right|_{g_{D^*D^*\pi} \to g_{D^*D^*\eta}},\nonumber\\
\mathcal{M}_{a5}^{mix}&= \left.\mathcal{M}_{a5}\dfrac{\sqrt 3}{4}\dfrac{m_d-m_u}{m_s-m}\right|_{g_{DD^*\pi} \to g_{DD^*\eta}},\nonumber\\
\mathcal{M}_{a5}^{mix}&= \left.\mathcal{M}_{a6}\dfrac{\sqrt 3}{4}\dfrac{m_d-m_u}{m_s-m}\right|_{g_{D^*D^*\pi} \to g_{D^*D^*\eta}},\nonumber
\end{array}
\end{equation}
with $m=(m_u+m_d)/2$.

Following our work~\cite{Qi:2025hwd} and the study of $\psi(3770)\to J/\psi\pi$~\cite{Guo:2012tj}, when calculating the Feynman amplitudes and decay width, a form factor
\begin{equation}
\begin{array}{rl}
\mathcal{F}(m_q^2,q^2)= \dfrac{m_q^2-\Lambda^2}{q^2-\Lambda^2}
\end{array}
\end{equation}
should be added to account for the off-shell effects of the effective coupling constants $g_{\psi D^{(*)}D^{(*)}}$ and $g_{D^{*}D\pi}$,
where the parameter $\Lambda$ can be reparametrized as $\Lambda=m_q+\alpha\Lambda_{QCD}$ with $m_q$ to be the mass of the exchanged
meson and $\Lambda_{QCD}=220$ MeV.
$\alpha$ is a free parameter and generally it should be of the order of unity, but its specific value cannot be determined from first principles. A commonly adopted method is to fix its value or range by fitting the experimental data of a known decay process~\cite{Tornqvist:1993vu,Tornqvist:1993ng,Locher:1993cc,Li:1996yn}.

Then the total amplitude for $\psi(3770)\to J/\psi+\pi$ is
\begin{equation}
\label{eq1}
\begin{array}{rl}
\mathcal{M}_{\psi_1\to J/\psi+\pi}= \displaystyle\sum^{6}_{i=1} 2\mathcal{M}_{ai} \mathcal{F}^2(m_q^2,q^2),
\end{array}
\end{equation}
where the factor $2$ in front of $\mathcal{M}_{ai}$ refers to the charge-conjugate contribution.
We also note that the $\mathcal{M}_{ai}$ that contains charged or neutral charmed meson loop contributions will cancel each other under the up-down quark symmetry limit.
However, since we study the isospin-violating decay $\psi(3770)\to J/\psi\pi$, we will discuss the corresponding effective coupling constants for charged or neutral charmed meson loops and give numerical values in the next section.
Then, the partial width of $\psi(3770)\to J/\psi\pi$ can be estimated through
\begin{equation}
\label{eq2}
\begin{array}{rl}
\Gamma(\psi(3770)\to J/\psi\pi)= \dfrac{1}{3}\dfrac{1}{8\pi}\dfrac{|\vec{p}|}{m_{\psi_1}^2}|\mathcal{M}_{\psi_1\to J/\psi\pi}|^2,
\end{array}
\end{equation}
where the factor $1/3$ comes from average over the spin of the initial state, and $|\vec{p}|$ denotes the magnitude of the three-momentum
of the final state in the initial rest frame.

In the same way, we can construct the amplitudes for $\psi(3770)\to h_c \pi$, $\psi(3823)/\psi(3842)\to J/\psi\pi$ and $h_c\pi$, which are collected in Appendix~\ref{Sec:AppA} and Appendix~\ref{Sec:AppB}.

\section{numerical result}

In this section, we will present the relevant coupling constants first. By fitted the experimental data, we estimate the decay $\psi(3770)\to J/\psi\pi$ and the range of the model parameter $\alpha$ can be obtained. Then we estimate the corresponding values of $\psi(3770)\to h_c\pi$, $\psi(3823)/\psi(3842)\to J/\psi\pi$ and $\psi(3823)/\psi(3842)\to h_c\pi$ with this $\alpha$ range.

\subsection{Coupling constant}


Firstly, we study the effect from effective coupling constant difference.
The $g_{\psi_1 D \bar D}$ is different for charged or neutral $D^{*}$ meson~\cite{Guo:2012tj,Wang:2011yh,Zhang:2009gy}.
Following Ref.~\cite{ParticleDataGroup:2024cfk}, we have $BR(\psi(3770)\to D^0 \bar D^0)_{exp}=52\%$ and $BR(\psi(3770)\to D^\pm  D^\mp)_{exp}=41\%$.
Thus one can obtain $g_{\psi_1 D^0 \bar D^0}$ and $g_{\psi_1 D^+ D^-}$ immediately.
However, the value of $m_{D^\pm}$ is too close to that of $m_{\psi_1}/2$ and this will lead to the threshold to affect the value.
So in our fitting we take $m_D=(m_{D^0}+m_{D^\pm})/2$ to avoid this point.
In this way, we have $g_{\psi_1 D^0 \bar D^0}=13.8$ and $g_{\psi_1 D^+ D^-}=12.2$.
As for another two scenarios, we take $g_{\psi_1 D  D}=(g_{\psi_1 D^0 \bar D^0} + g_{\psi_1 D^+ D^-})/2$ for simplification.
While for other effective constants such as $g_{\psi_1 D^{(*)} D^*}$ and $g_{J/\psi D^{(*)} D^{(*)}}$, since those decays are under the threshold, we consider the values for neutral and charge are the same for they only contain off shell effect.

Another coupling constant is $g_{D^*D\pi}$. Although the experiment give concrete value as $BR(D^{*0}\to D^0 \pi^0)_{exp}=64.7\%$ and $BR(D^{*\pm}\to D^\pm \pi^0)_{exp}=30.7\%$, the exactly value of the total decay widths for $D^{*0}$ is not measured.
Thus in this study, we consider the values of $g_{D^*D\pi}$ for neutral and charge are the same.
As for charmed-strange meson loop, even though the experiment give $BR(D^{*\pm}_s\to D^+_s \pi^0)_{exp}=5.77\%$, but the total decay width for $D^{*}_s$ is lacked.
Considering this point and the ideal mixing of $\pi-\eta-\eta'$~\cite{Casalbuoni:1996pg,Casalbuoni:1992gi,Casalbuoni:1992dx}, we do not consider the charmed-strange meson loop in our study.

As for the charged and neutral charmed meson mass difference in meson loop, we take $m_{D^0}=1.865$ GeV and $m_{D^{*0}}=2.007$ GeV for neutral meson loop and $m_{D^\pm}=1.870$ GeV and $m_{D^{*\pm}}=2.010$ GeV for charge meson loop respectively.
In another two scenarios, we take $m_{D}=1.865$ GeV and $m_{D^{*}}=2.007$ GeV for simplification.

The third way is $\eta-\pi$ mixing contribution.
Following Refs.~\cite{Qi:2025xkz,Qi:2025hwd,Xiao:2016hoa,Yue:2025wcl}, we can give the concrete expression for  the corresponding Feynman diagrams listed in Fig.~\ref{fig1mix}.

In our calculation, all of the other effective coupling constants relevant to D-wave charmonia and charmed mesons are listed as~\cite{Qi:2023kwc}
\begin{eqnarray}
	g_{\psi_1 DD}&=&-2g_2 \frac{\sqrt{15}}{3}\sqrt{m_{\psi_1}m_D m_D}\ ,\nonumber\\
	g_{\psi_1 D^*D}&=&g_2 \frac{\sqrt{15}}{3}\sqrt{m_{D}m_{D^\ast}/m_{\psi_1}}\ ,\nonumber\\
	g_{\psi_1 D^* D^*}&=&-g_2 \frac{\sqrt{15}}{15}\sqrt{m_{\psi_1}m_{D^\ast} m_{D^\ast}}\ ,\nonumber\\
	g_{\psi_2 D^* D}&=&2g_2 \sqrt{\frac{3}{2}}\sqrt{m_{\psi_2}m_D m_{D^\ast}}\ ,\nonumber\\
	g_{\psi_2 D^* D^*}&=&-2g_2 \sqrt{\frac{1}{6}}\sqrt{m_{D^\ast}m_{D^\ast}/m_{\psi_2}}\ ,\nonumber\\
	g_{\psi_3 D^* D^*}&=&2g_2 \sqrt{m_{\psi_3}m_{D^\ast} m_{D^\ast}}\ ,
\end{eqnarray}
with $g_{\psi_1 D D}=13.0$ as mentioned above.

For the coupling constants relevant to P-wave and S-wave charmonia and charmed mesons,  we just adopt VMD model and list as~\cite{Colangelo:2003sa,Falk:1992cx,Chen:2014sra,Yan:1992gz,Cheng:1992xi,Wise:1992hn}
\begin{eqnarray}
	g_{h_c D^*D}&=&-2g_3 \sqrt{m_{h_c} m_{D}m_{D^\ast}}\ ,\nonumber\\
	g_{h_c D^* D^*}&=&2g_3 \sqrt{ \frac{m_{D^\ast}^2}{m_{h_c}} }\ ,\nonumber\\
	g_{\psi D D} &=&2 g_{1} \sqrt{m_{\psi}} m_{D}\  ,\nonumber\\
	g_{\psi D^{*} D} &=&2 g_{1} \sqrt{m_{D} m_{D^{*}} / m_{\psi}}\ ,\nonumber \\
	g_{\psi D^{*} D^{*}} &=&2 g_{1} \sqrt{m_{\psi}} m_{D^{*}}\ ,
\end{eqnarray}
where the gauge coupling is defined as $g_1= \sqrt{m_{\psi}}/\left(2m_Df_\psi \right)$ with $f_\psi=416$ MeV to be the decay constant of $J/\psi$, which can be estimated from the dilepton decay width of $J/\psi$~\cite{Colangelo:2003sa,ParticleDataGroup:2024cfk}.
The parameter $g_3$ is defined as
\begin{eqnarray}
g_3=-\frac{m_{\chi_{c0}} }{3}\frac{1}{f_{\chi_{c0}} }\ ,\nonumber\\
\end{eqnarray}
with $f_{\chi_{c0}}=510$ MeV~\cite{Colangelo:2003sa,Wu:2025crk}.

As for the coupling constants related to the charmed meson and  light pseudoscalar, under the heavy quark limit and chiral symmetry~\cite{Falk:1992cx,Chen:2014sra,Yan:1992gz,Cheng:1992xi,Wise:1992hn}, we have
\begin{eqnarray}
	g_{D^{\ast}D\eta}&=&\alpha' \frac{2g}{f_\pi}\sqrt{m_{D^\ast}m_{D}}\ ,\nonumber\\
	g_{D^{\ast}D^\ast \eta }&=&\alpha' \frac{2g}{f_\pi}\ ,\nonumber\\
	g_{D^{\ast}D\pi}&=&\frac{1}{\sqrt2} \frac{2g}{f_\pi}\sqrt{m_{D^\ast}m_{D}}\ ,\nonumber\\
	g_{D^{\ast}D^\ast \pi}&=&\frac{1}{\sqrt2} \frac{2g}{f_\pi}\ ,
\end{eqnarray}
where $f_\pi=132\ \mathrm{MeV}$ is the decay constant of pion, and $g=0.59$, which is evaluated by the partial width of $D^\ast\to D\pi$~\cite{ParticleDataGroup:2024cfk}.

\subsection{Decays of $\psi(3770)$ to $J/\psi\pi$  and $h_c\pi$ }

\begin{figure}[t]
\includegraphics[width=8.2cm]{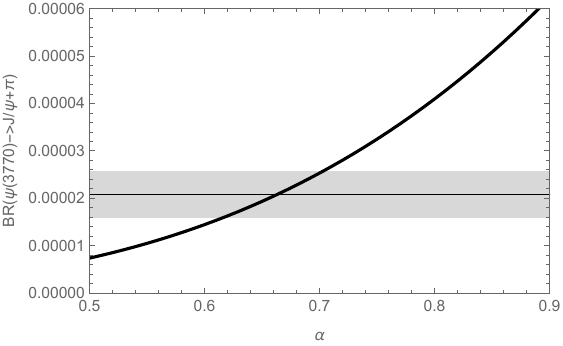}
\caption{The branching ratio of $\psi(3770)\to J/\psi\pi$ depending on the model parameter $\alpha$.
The gray banded area with thin line stands for the experimental measured branching
fraction of with $BR(\psi(3770)\to J/\psi\pi)=(2.08\pm0.49)\times10^{-5}$.
The area in this region stands for the region allowed by experiment.
\label{3770value}}
\end{figure}

Under the above preparations, we can calculate the branching ratios of the relevant hidden charm decay processes depending on the model parameter $\alpha$.
Then we present the numerical result of $\psi(3770)\to J/\psi\pi$ depending on the model parameter $\alpha$ in Fig.~\ref{3770value}.
From this figure, one can find the branching ratio increase as the increasing of the parameter $\alpha$.
In addition, Since the branching ratio of $\psi(3770)\to J/\psi+\pi$ has been measured to be $(2.08\pm0.49)\times10^{-5}$~\cite{BESIII:2026pyf}, which corresponds to $\alpha=0.66\pm0.05$ in Fig.~\ref{3770value}.

Then we calculate the three contributions to affect the isospin violated channel separately and list the corresponding numerical values in Table \ref{tab1}.
In this table we take $\alpha=0.61\sim0.71$ as the fitted parameters space from experiment.
\begin{table*}[htbp]
\begin{center}
\caption{\label{tab1} The branching ratios of $\psi(3770)\to J/\psi\pi$ under three scenarios, the $\delta g_{\psi_1DD}\neq0$, $\delta m_{D^{(*)}}\neq0$ and $\eta-\pi$ mixing, respectively. $(a_i)$ corresponds the Feynman diagram listed in Fig.~\ref{fig1}.
}
\renewcommand\arraystretch{1.5}
\begin{tabular*}{160mm}{c@{\extracolsep{\fill}}cccccccc}
\toprule[1pt]
  Scenarios    &  $(a1)$     &      $(a2 - a4)$                &      $(a5, a6)$    & All diagrams \\
\midrule[1pt]
$\delta g_{\psi_1DD}\neq0$ &  $(6.07\sim9.81)\times10^{-6}$ & -                     &   -   &    $(6.07\sim9.81)\times10^{-6}$ \\
$\delta m_{D^{(*)}}\neq0$  & $(2.51\sim4.05)\times10^{-7}$  & $(1.53\sim2.47)\times10^{-7}$ & $(1.27\sim2.05)\times10^{-8}$ &  $(1.01\sim1.63)\times10^{-6}$   \\
$\eta-\pi$ mixing        & $(3.69\sim5.96)\times10^{-8}$   & $(5.74\sim9.28)\times10^{-8}$  & $(9.91\sim16.02)\times10^{-9}$  & $(2.82\sim4.56)\times10^{-7}$ \\
\midrule[1pt]
Total                 &  $(1.00\sim1.61)\times10^{-5}$ & $(3.98\sim6.43)\times10^{-7}$ & $(4.50\sim7.28)\times10^{-8}$ &    $(1.60\sim2.59)\times10^{-5}$ \\
\bottomrule[1pt]

\end{tabular*}
\vspace{0mm}
\end{center}
\end{table*}
In Table~\ref{tab1}, we present the individual contributions from different meson loops.

From the last line of  this table, we can find that contribution from diagram $(a1)$ is two orders larger than that from diagrams $(a2 - a4)$, and their contribution are $2\sim3$ orders larger than that from diagrams $(a5, a6)$.
It also show that even if only considering the diagram containing $(a1)$, the contribution from $\delta g_{\psi_1 DD}\neq0$ frame is $1\sim 2$ orders larger than that from $\delta m_{D^{(*)}}\neq0$ scenario.
After considering all the diagrams, the contribution from $\delta g_{\psi_1DD}\neq0$ scenario is one orders larger than that from $\delta m_{D^{(*)}}\neq0$ scenario.

As for the $\eta-\pi$ mixing contribution, its contribution is one order smaller than that from $\delta m_{D^{(*)}}\neq0$ scenario, and two orders smaller than that from $\delta g_{\psi_1 DD}\neq0$ scenario.
It means that the contribution from onshell meson loop contribute the most, and contribution from $\eta-\pi$ mixing is non-negligible for excited D-wave charmonia states.
With this fixed model parameter range of $\alpha$, we could further estimate the branching fractions of other involved decay processes.

\begin{figure}[t]
\includegraphics[width=8.2cm]{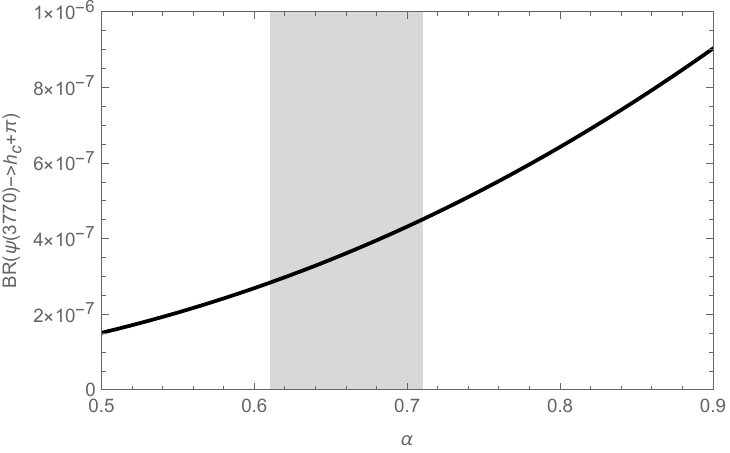}
\caption{The branching ratio of $\psi(3770)\to h_c\pi$  depending on the model parameter $\alpha$.
The gray banded area stands for the range with $\alpha=0.61\sim0.71$.
\label{3770hcvalue}}
\end{figure}

Under the parameter $\alpha=0.61\sim0.71$ fitted above through $BR(\psi(3770)\to J/\psi\pi)_{exp}$, we estimate the branching ratio of  $\psi(3770)\to h_c\pi$.
The corresponding Feynman diagrams and amplitudes are listed in Appendix \ref{Sec:AppA}.
After calculation, we list the numerical result in Fig.~\ref{3770hcvalue}.
With the parameter range fitted through $\psi(3770)\to J/\psi\pi$, we predict $BR(\psi(3770)\to h_c\pi)=(2.7\sim4.4)\times10^{-7}$.
This value is about two orders smaller than that from $\psi(3770)\to J/\psi\pi$.
We consider that the tiny phase space of $h_c\pi$ highly suppress the numerical value of $BR(\psi(3770)\to h_c\pi)$.

\subsection{Decays of $\psi(3823)/\psi(3842)$ to $J/\psi\pi$ and $h_c\pi$}

With the above parameter range, we can study the decays of $\psi(3823)$ and $\psi(3842)$ to $J/\psi\pi$ and $h_c\pi$, then  estimate their decay widths respectively.
Based on the effective Lagrangians above and following our pervious study, we have five Feynman diagrams for $\psi(3823)\to J/\psi\pi$, two diagrams for $\psi(3842)\to J/\psi\pi$,  four diagrams for $\psi(3823)\to h_c\pi$, and two diagrams for $\psi(3842)\to h_c\pi$, respectively.
All of the corresponding Feynman diagrams and Feynman amplitudes are listed in Appendix \ref{Sec:AppB}.

From Appendix \ref{Sec:AppB}, one can find that the coupling $\psi(3823)D\bar D$ and $\psi(3842)D\bar D$ in the meson loops are not emerge in our calculation, since $\psi(3823)D\bar D$ is forbidden by the conservations of $J^P$ quantum numbers, and $\psi(3842)D\bar D$ vanish in the leading order of the effective Lagrangian obtained from the expansion of Eq.~(\ref{Lagrangian}).
Thus, in the the present estimations the contribution comes from onshell loop will not be considered.

\begin{figure}[t]
\includegraphics[width=8.2cm]{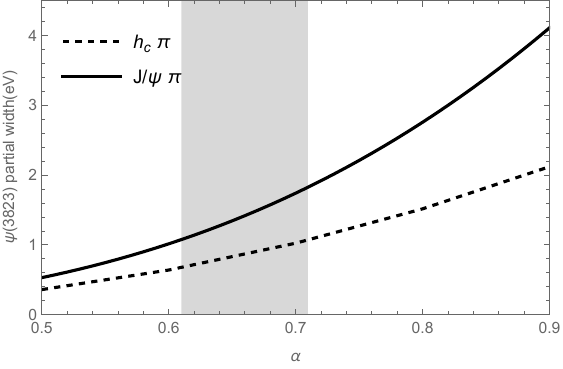}
\caption{The partial widths of $\psi(3823)$ to $J/\psi\pi$ and $h_c\pi$ depending on the model parameter $\alpha$.
The gray banded area stands for the range with $\alpha=0.61\sim0.71$.
\label{3823value}}
\end{figure}

Since the precise total decay width of $\psi(3823)$ is not clearly so far by now and the experiment only give the upper limit $\Gamma_{\psi(3823)}<2.9$ MeV~\cite{ParticleDataGroup:2024cfk}.
Therefore, we estimate the decay widths of them instead of exploring the branching fractions and list the partial widths of $\psi(3823)$ to $J/\psi\pi$ and $h_c\pi$ in Fig.~\ref{3823value}.
With the parameter range determined through $\psi(3770)\to J/\psi\pi$, we have
\begin{equation}
\begin{array}{rl}
&\Gamma(\psi(3823)\to J/\psi\pi)=(1.12\sim1.77) ~\mathrm{eV},\\
&\Gamma(\psi(3823)\to h_c\pi)=(0.71\sim1.04) ~\mathrm{eV}.
\end{array}
\end{equation}
In addition, we note that the values of the partial width of these two hidden charm decay channels $\psi(3823)$ to $J/\psi\pi$ and $h_c\pi$ are of the same orders, which is different from the circumstance of $\psi(3770)$.
We consider it is caused through the larger phase space of $\psi(3823)$ than the $\psi(3770)$ for $h_c\pi$ final state.

As for $\psi(3823)$ decay channels, since the coupling between $\psi(3823)$ and $D\bar D$ is not exist, the dominant contribution to isospin violated channels come from $\delta m_{D^{(*)}}\neq0$, and contribution from $\eta-\pi$ mixing is non-negligible.

\begin{figure}[t]
\includegraphics[width=8.2cm]{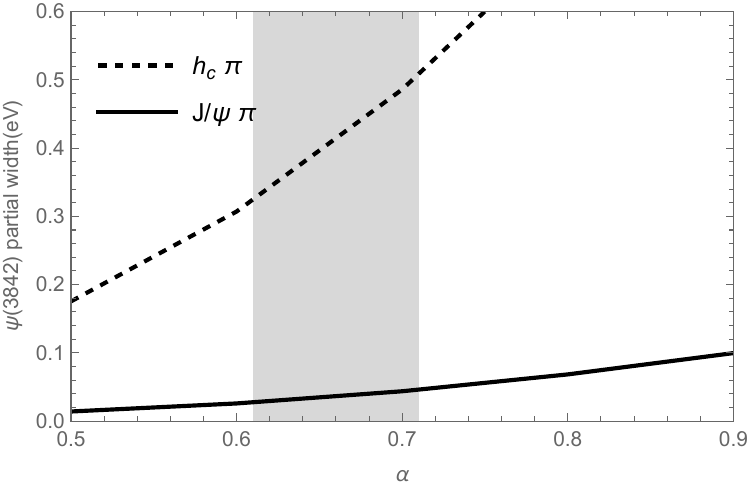}
\caption{The same as Fig.~\ref{3823value} but for $\psi(3842)\to J/\psi\pi$ and $h_c\pi$.
\label{3842value}}
\end{figure}

In a same manner, we can evaluate the hidden charm decays of $\psi(3842)$ and estimate the partial decay widths of $\psi(3842)$ to $J/\psi\pi$ and $h_c\pi$, which are presented in Fig.~\ref{3842value}.
From the figure, we can read
\begin{equation}
\begin{array}{rl}
&\Gamma(\psi(3842)\to J/\psi \pi)=(0.028\sim0.044) ~\mathrm{eV},\\
&\Gamma(\psi(3842)\to h_c \pi)=(0.34\sim0.49) ~\mathrm{eV}.
\end{array}
\end{equation}

We also list the branching fractions of the light meson decays of $\psi(3823)$ and $\psi(3842)$ resulted from different meson loops in Table \ref{tab2}.
\begin{table*}[htbp]
\begin{center}
\caption{\label{tab2}
The partial decay widths of $\psi(3823)$ and $\psi(3842)$ resulted from different meson loops and the corresponding Feynman diagrams are listed in Appendix \ref{Sec:AppB}.
The decay width are in unit of eV.
}
\renewcommand\arraystretch{1.5}
\begin{tabular*}{160mm}{c@{\extracolsep{\fill}}ccccc}
\toprule[1pt]
  Channel          &  Scenario            &  $(b1-b3)/(e1,e2)$     &      $(b4,b5)/(e3,e4)$   &      $(c1,c2)/(f1,f2)$ &       All diagrams \\
\midrule[1pt]
$\psi_2\to J/\psi \pi$ & $\delta m_{D^{(*)}}\neq0$ & $(0.21\sim0.33)$ & $(0.032\sim0.051)$ &   -   &   $(0.41\sim0.64)$ \\
                        &  $\eta-\pi$ mixing       &  $(0.042\sim0.069)$ & $(0.044\sim0.072)$ &  -  &   $(0.17\sim0.28)$ \\
$\psi_2\to h_c \pi$  &  $\delta m_{D^{(*)}}\neq0$ & $(0.34\sim0.50)$ &  $(0.008\sim0.012)$ &  -      &  $(0.46\sim0.67)$  \\
                      &  $\eta-\pi$ mixing         & $(0.019\sim0.028)$ &  $(0.0008\sim0.0013)$ &   -  &  $(0.027\sim0.041)$   \\
$\psi_3\to J/\psi \pi$ & $\delta m_{D^{(*)}}\neq0$ &              -   &       -       & $(0.015\sim0.023)$ & $(0.015\sim0.023)$ \\
                        &  $\eta-\pi$ mixing        &               -   &     -    &$(0.0022\sim0.0034)$ & $(0.0022\sim0.0034)$ \\
$\psi_3\to h_c \pi$    & $\delta m_{D^{(*)}}\neq0$ &  -              &  -         & $(0.20\sim0.29)$ &    $(0.20\sim0.29)$ \\
                        &  $\eta-\pi$ mixing        &  -                 &  -         & $(0.017\sim0.026)$ &    $(0.017\sim0.026)$ \\
\bottomrule[1pt]

\end{tabular*}
\vspace{0mm}
\end{center}
\end{table*}
The circumstance of $\psi(3842)$ are the same as that of $\psi(3823)$, namely the contribution form onshell meson D loop is not emerged.
Moreover, from Table \ref{tab2} we can find that the contribution from $g_{\psi_2 DD^*}$ is about one order larger than that from $g_{\psi_2 D^*D^*}$.
This conclusion is coincide with the conclusion from Table \ref{tab1}.

We also note that, since the on-shell charmed meson loop is not existed or emerged, the contribution from $\eta-\pi$ mixing frame become non-neglectable.
As for $\psi(3842)$ decay channels, since the $g_{\psi_3 DD^*}$ is lacked, the fractions of $\psi(3842)$ decay channels is one order smaller than that of $\psi(3823)$.
Table \ref{tab2} also show that, as the mass of $\psi_i$ increase and the momentum of the final state meson becomes larger, the decay width of $h_c\pi$ final state is larger than that of $J/\psi\pi$.

\section{Summary}

The experimental observations of $\psi(3823)$ and $\psi(3842)$ enrich the D-wave charmonia family.
In our previous study we calculate the isospin conserved hadronic decays of three D-wave charmonia $\psi(3770)$, $\psi(3823)$, and $\psi(3842)$ to $J/\psi\eta$ and $\eta_c\omega$.
Different from isospin conserved decays, the isospin-violating decays are vanished under $SU(3)$ limit for charged and neutral charmed meson loops cancel each other and the charmed-strange mesons have no effective coupling constant with $\pi$.

Very recently, the experiment observed the isospin violated decay channel and measured its branching ratio as $BR(\psi(3770)\to J/\psi \pi)=(2.08\pm0.49)\times10^{-5}$~\cite{BESIII:2026pyf}.
Thus in this work we study the isospin-violating decays $\psi(3770)\to h_c \pi$, $\psi(3823)/\psi(3842)$ decay to $J/\psi \pi$ and $h_c \pi$ through charmed meson loops mechanism respectively,
which is believe to be the phenomenological description of the long-range distributions in higher D-wave charmonia decay processes.
The model parameter $\alpha$ is fitted by reproducing the branching fraction of $\psi(3770)\to J/\psi \pi$.

Our work are under the three scenarios which contribute to the isospin violation and are listed in Table \ref{tab1}.
After calculation, we find that for $\psi(3770)$ decay channels, the on shell contribution from $\delta g_{\psi_1DD}\neq0$ scenario offer the most part, and contribution from $\delta m_{D^{(*)}}\neq0$ scenario are one order smaller than that from $\delta g_{\psi_1DD}\neq0$ scenario.
The contribution from $\eta-\pi$ mixing scenario is more than two orders smaller than the two scenarios above and can be neglected as it was done in Ref.~\cite{Guo:2012tj}.

As for $\psi(3823)/\psi(3842)$ decay channels, since the onshell meson loops was not contained in the present estimations for this vertex is forbidden or vanish in the leading order of the effective Lagrangian.
The contribution from $\eta-\pi$ mixing are just one order smaller than that from $\delta m_{D^{(*)}}\neq0$ scenario, it means that for higher excited charmonium decays, the contribution from $\eta-\pi$ mixing should be considered if we do not consider the onshell meson loop contribution to the decay channels.
On the another aspect, it means that the values listed in Figs.~\ref{3823value} and~\ref{3842value} represent the lower limit of branching ratios and decay widths, since in those calculations we do not contain the onshell charmed meson loop contributions.
And as it knows to all, the contribution from effective coupling constants usually appears more dominant in branching ratios and decay widths.

Moreover, in our study the difference between $g_{D^{*\pm}D^{\pm}\pi}$ and $g_{D^{*0}D^*\pi}$ are not considered.
Unfortunately, although the branching ratios of $D^*\to D \pi$ for $D^0$, $D^\pm$ and $D_s$ are measured clearly, but the total decay widths of $D^{*0}$ and $D^*_s$ are not measured in experiment clearly.
Thus in our present work we do not consider neither the tinny difference between them nor the contribution from $g_{D_s^{*\pm}D_s^{\pm}\pi}$ to meson loop contribution.
We expect more accurate measurement about $D^{*0}$ and $D^*_s$ will improve the research on charmonia isospin violation decays.
Moreover, we hope more precise measurements from the BES III, Belle II, and LHCb Collaborations could further test the corresponding decay widths and the results in the present work.

\section*{Acknowledgments}

This work is supported by the National Natural Science Foundation of China under the Grants No.12405093 as well as supported, in part, by the National Key Re-search and Development Program under Grants No. 2024YFA1610504.


\appendix

\section{Amplitudes for $\psi(3770)\to h_c \pi$\label{Sec:AppA}}

\begin{figure}[htbp]
\includegraphics[width=8.2cm]{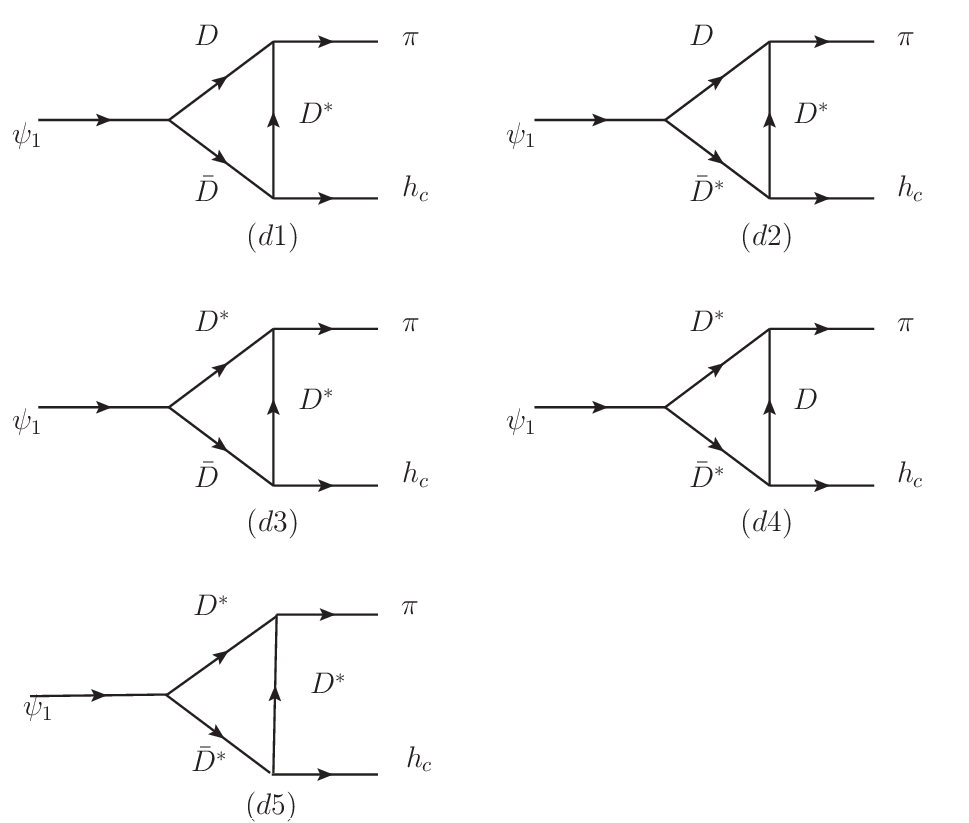}
\caption{Feynman diagrams contributing to $\psi(3770)\to h_c\pi$ at the hadron level.\label{figa}}
\end{figure}

The meson loops contribute to $\psi(3770)\to h_c \pi$ are listed in Fig.~\ref{figa}.
The amplitudes corresponding to Fig.~\ref{figa} read,

\begin{equation}
\begin{array}{rl}
\mathcal{M}_{d1}= &i^3\displaystyle\int \dfrac{d^4 q}{(2\pi)^4}[-i g_{\psi_1 DD}(-i)(-1)(p_1-p_2)\cdot\epsilon_{\psi_1}(p)]\\
& [-i g_{D^*D\pi}(-ip_3)^\alpha][i g_{h_c D^*D} \epsilon_{h_c}^\beta(p_4)]\\
& \dfrac{i}{p_1^2-m_D^2}\dfrac{i}{p_2^2-m_D^2}\dfrac{-g^{\alpha\beta}+q^\alpha q^\beta/m_{D^*}^2}{q^2-m_{D^*}^2}
\end{array}
\end{equation}

\begin{equation}
\begin{array}{rl}
\mathcal{M}_{d2}= &i^3\displaystyle\int \dfrac{d^4 q}{(2\pi)^4}
[g_{\psi_1 D^*D} \epsilon_{\mu\nu\alpha\kappa} (-ip)^\mu  \epsilon_{\psi_1}^\nu(p) (-i)(p_1-p_2)^\alpha]\\
& [-i g_{D^*D\pi}(-i p_3)^\tau] [ g_{h_c D^* D^*}\epsilon_{\rho\sigma\chi\beta}(-ip_4)^\rho \epsilon_{h_c}^\sigma(p_4) ]\\
& \dfrac{1}{p_1^2-m_D^2}\dfrac{-g^{\kappa\beta}+p_2^\kappa p_2^\beta/m_{D^*}^2}{p_2^2-m_{D^*}^2}
\dfrac{-g^{\tau\chi}+q^\tau q^\chi/m_{D^*}^2}{q^2-m_{D^*}^2}
\end{array}
\end{equation}

\begin{equation}
\begin{array}{rl}
\mathcal{M}_{d3}= &i^3\displaystyle\int \dfrac{d^4 q}{(2\pi)^4}
[g_{\psi_1 D^*D} \epsilon_{\mu\nu\alpha\beta}(-1) (-ip)^\mu  \epsilon_{\psi_1}^\nu(p)\\
& (-i)(p_1-p_2)^\alpha][\dfrac{g_{D^*D^*\pi} }{2}
\epsilon_{\chi\lambda\sigma\rho}(-1)(-i p_3)^\lambda(-i)(p_1+q)^\sigma]\\
& [i g_{h_c D^*D} \epsilon_{h_c}^\tau(p_4) ]\dfrac{-g^{\rho\beta}+p_1^\rho p_1^\beta/m_{D^*}^2}{p_1^2-m_{D^*}^2}
\dfrac{1}{p_2^2-m_D^2}\\
&\dfrac{-g^{\chi\tau}+q^\chi q^\tau/m_{D^*}^2}{q^2-m_{D^*}^2}
\end{array}
\end{equation}

\begin{equation}
\begin{array}{rl}
\mathcal{M}_{d4}= &i^3\displaystyle\int \dfrac{d^4 q}{(2\pi)^4}
[i g_{\psi_1 D^*D^*}(-i)(-4\epsilon_{\psi_1}(p)\cdot (p_1-p_2)g^{\alpha\beta} \\
&+\epsilon_{\psi_1}^\beta(p)(p_1-p_2)^\alpha +\epsilon_{\psi_1}^\alpha(p)(p_1-p_2)^\beta)]\\
& [-i g_{D^*D\pi}(-1)(-i p_3)^\mu]  [i g_{h_c D^*D} \epsilon_{h_c}^\nu(p_4) ]\\
&\dfrac{-g^{\mu\alpha}+p_1^\mu p_1^\alpha/m_{D^*}^2}{p_1^2-m_{D^*}^2}
\dfrac{-g^{\nu\beta}+p_2^\nu p_2^\beta/m_{D^*}^2}{p_2^2-m_{D^*}^2} \dfrac{1}{q^2-m_D^2}
\end{array}
\end{equation}

\begin{equation}
\begin{array}{rl}
\mathcal{M}_{d5}= &i^3\displaystyle\int \dfrac{d^4 q}{(2\pi)^4}  [i g_{\psi_1 D^*D^*}(-i)
(-4\epsilon_{\psi_1}(p)\cdot (p_1-p_2)g^{\alpha\beta}\\
&+\epsilon_{\psi_1}^\beta(p)(p_1-p_2)^\alpha +\epsilon_{\psi_1}^\alpha(p)(p_1-p_2)^\beta)]\\
&[\dfrac{g_{D^*D^*\pi} }{2}\epsilon_{\chi\lambda\sigma\rho}(-1)(-i p_3)^\lambda(-i)(p_1+q)^\sigma] \\
& [ g_{h_c D^* D^*}\epsilon_{\kappa\tau\mu\nu}(-ip_4)^\kappa \epsilon_{h_c}^\tau(p_4) ]\\
&\dfrac{-g^{\rho\beta}+p_1^\rho p_1^\beta/m_{D^*}^2}{p_1^2-m_{D^*}^2}
\dfrac{-g^{\mu\chi}+p_2^\mu p_2^\chi/m_{D^*}^2}{p_2^2-m_{D^*}^2}
\dfrac{-g^{\nu\alpha}+q^\nu q^\alpha/m_{D^*}^2}{q^2-m_{D^*}^2}
\end{array}
\end{equation}

Then the total amplitude for $\psi(3770)\to h_c \pi$ reads
\begin{equation}
\label{eq1}
\begin{array}{rl}
\mathcal{M}_{\psi(3770)\to h_c\pi}= \displaystyle\sum^{5}_{i=1} 2\mathcal{M}_{di} \mathcal{F}^2(m_q^2,q^2)
\end{array}
\end{equation}
with the factor $2$ in front of $\mathcal{M}_{di}$ refers to the charge conjugate contribution.

The amplitudes contributing to the $\psi(3770)\to h_c\pi$ process through $\eta-\pi$ mixing, can be obtained by employing the following substitutions
\begin{equation}
\begin{array}{rl}
\mathcal{M}_{\psi(3770)\to h_c\pi}^{mix}&= \left.\mathcal{M}_{\psi(3770)\to h_c\eta}\dfrac{\sqrt 3}{4}\dfrac{m_d-m_u}{m_s-m}\right|_{g_{DD^*\pi} \to g_{DD^*\eta}},\nonumber\\
\label{mix}
\end{array}
\end{equation}
with $m=(m_u+m_d)/2$.

\section{Amplitudes for $\psi(3823)/\psi(3842)\to J/\psi \pi$\label{Sec:AppB}}

\begin{figure}[htbp]
\includegraphics[width=8.2cm]{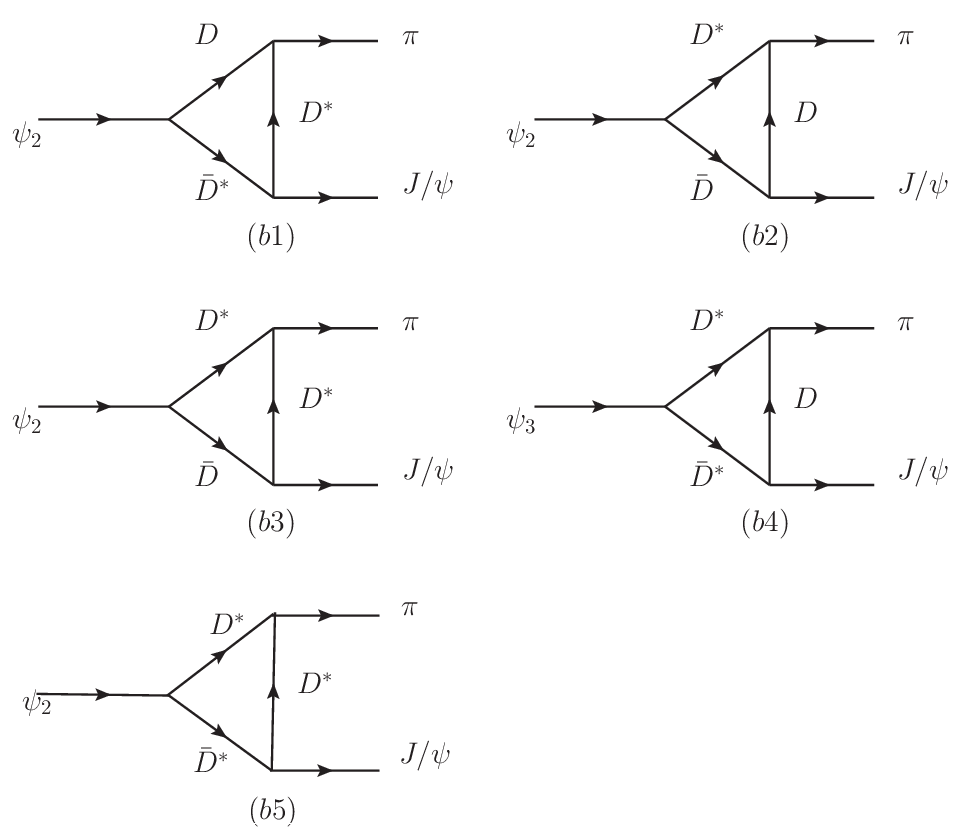}
\caption{Feynman diagrams contributing to $\psi(3823)\to J/\psi\pi$ at the hadron level.\label{figb1}}
\end{figure}

The meson loops contribute to $\psi(3823)\to J/\psi \pi$ are listed in Fig.~\ref{figb1}.
The amplitudes corresponding to Fig.~\ref{figb1} read,

\begin{equation}
\begin{array}{rl}
\mathcal{M}_{b1}= &i^3\displaystyle\int \dfrac{d^4 q}{(2\pi)^4}
[i g_{\psi_2 D^*D}   \epsilon_{\psi_2}^{\mu\alpha}(p)(-1)(-i)(p_1-p_2)^\mu]\\
&[-i g_{D^*D\pi}(-i p_3)^\beta] (i g_{\psi D^* D^*} )\\
&[(-1)(-i)(-\epsilon_\psi(p_4)\cdot (q-p_2)g^{\nu\chi} \\
&+\epsilon_\psi^\nu(p_4)(q-p_2)^\chi +\epsilon_\psi^\chi(p_4)(q-p_2)^\nu)]\\
&\dfrac{1}{p_1^2-m_D^2} \dfrac{-g^{\nu\alpha}+p_2^\nu p_2^\alpha/m_{D^*}^2}{p_2^2-m_{D^*}^2} \\
&\dfrac{-g^{\beta\chi}+ q^\beta q^\chi/m_{D^*}^2}{q^2-m_{D^*}^2}
\end{array}
\end{equation}

\begin{equation}
\begin{array}{rl}
\mathcal{M}_{b2}= &i^3\displaystyle\int \dfrac{d^4 q}{(2\pi)^4}
[i g_{\psi_2 D^*D} \epsilon_{\psi_2}^{\mu\alpha}(p) (-i)(p_1-p_2)^\mu]\\
& [-i g_{D^*D\pi}(-1)(-i p_3)^\nu][-i g_{\psi DD}(-i)(q-p_2)\cdot\epsilon_\psi(p_4)]\\
& \dfrac{-g^{\nu\alpha}+p_1^\nu p_1^\alpha/m_{D^*}^2}{p_1^2-m_{D^*}^2}\dfrac{1}{p_2^2-m_D^2}
\dfrac{1}{q^2-m_D^2}
\end{array}
\end{equation}

\begin{equation}
\begin{array}{rl}
\mathcal{M}_{b3}= &i^3\displaystyle\int \dfrac{d^4 q}{(2\pi)^4}
[i g_{\psi_2 D^*D} \epsilon_{\psi_2}^{\mu\alpha}(p) (-i)(p_1-p_2)^\mu]\\
& [\dfrac{g_{D^*D^*\pi} }{2} \epsilon_{\chi\lambda\sigma\nu}(-1)(-i p_3)^\lambda(-i)(p_1+q)^\sigma] \\
& [g_{\psi D^*D} \epsilon_{\kappa\tau\phi\xi} (-i p_4)^\kappa \epsilon_\psi^\tau(p_4)(-i)(q-p_2)^\phi ]\\
&\dfrac{-g^{\nu\alpha}+p_1^\nu p_1^\alpha/m_{D^*}^2}{p_1^2-m_{D^*}^2}\dfrac{1}{p_2^2-m_D^2} \\
&\dfrac{-g^{\xi\chi}+q^\xi q^\chi/m_{D^*}^2}{q^2-m_{D^*}^2}
\end{array}
\end{equation}

\begin{equation}
\begin{array}{rl}
\mathcal{M}_{b4}= &i^3\displaystyle\int \dfrac{d^4 q}{(2\pi)^4}
[g_{\psi_2 D^*D^*}\epsilon_{\lambda\nu\alpha\beta}(-1)(-i p^\lambda)\epsilon_{\psi_2}^{\mu\nu}(p)\\
&(-i)(g^{\alpha\sigma}(p_1-p_2)^\beta g^{\mu\delta} +g^{\mu\sigma}(p_2-p_1)^\beta g^{\alpha\delta})]\\
& [-i g_{D^*D\pi}(-1)(-i p_3)^\theta] \\
& [g_{\psi D^*D} \epsilon_{\kappa\tau\phi\xi} (-i p_4)^\kappa \epsilon_\psi^\tau(p_4) (-i)(-1)(q-p_2)^\phi ]\\
&\dfrac{-g^{\sigma\theta}+p_1^\sigma p_1^\theta/m_{D^*}^2}{p_1^2-m_{D^*}^2}
\dfrac{-g^{\xi\delta}+p_2^\xi p_2^\delta/m_{D^*}^2}{p_2^2-m_{D^*}^2} \\
& \dfrac{1}{q^2-m_D^2}
\end{array}
\end{equation}

\begin{equation}
\begin{array}{rl}
\mathcal{M}_{b5}= &i^3\displaystyle\int \dfrac{d^4 q}{(2\pi)^4}
[g_{\psi_2 D^*D^*}\epsilon_{\lambda\nu\alpha\beta}(-1)(-i p^\lambda)\epsilon_{\psi_2}^{\mu\nu}(p)\\
&(-i)(g^{\alpha\sigma}(p_1-p_2)^\beta g^{\mu\kappa} +g^{\mu\sigma}(p_2-p_1)^\beta g^{\alpha\kappa})]\\
& [\dfrac{g_{D^*D^*\pi} }{2}\epsilon_{\theta\tau\phi\rho}(-1)(-i p_3)^\tau(-i)(p_1+q)^\phi ] \\
& [i g_{\psi D^*D^*}(-i)(-1)(-\epsilon_{\psi_1}(p_4)\cdot (q-p_2)g^{\delta\chi}\\
&+\epsilon_\psi^\chi(p_4)(q-p_2)^\delta +\epsilon_\psi^\delta(p_4)(q-p_2)^\chi)]\\
& \dfrac{-g^{\sigma\rho}+ p_1^\sigma p_1^\rho/m_{D^*}^2}{p_1^2-m_{D^*}^2}
\dfrac{-g^{\delta\kappa}+ p_2^\delta p_2^\kappa/m_{D^*}^2}{p_2^2-m_{D^*}^2} \\
&\dfrac{-g^{\chi\theta}+ q^\chi q^\theta/m_{D^*}^2}{q^2-m_{D^*}^2}
\end{array}
\end{equation}

Then the total amplitude for $\psi(3823)\to J/\psi \pi$ read,
\begin{equation}
\label{eq1}
\begin{array}{rl}
\mathcal{M}_{\psi(3823)\to J/\psi\pi}= \displaystyle\sum^{5}_{i=1} 2\mathcal{M}_{bi} \mathcal{F}^2(m_q^2,q^2)
\end{array}
\end{equation}
with the factor $2$ in front of $\mathcal{M}_{bi}$ refers to the charge conjugate contribution. The amplitudes contributing to the $\psi(3823)\to J/\psi\pi$ process through $\eta-\pi$ mixing can be obtained through Eq.~(\ref{mix}).

\begin{figure}[htbp]
\includegraphics[width=8.2cm]{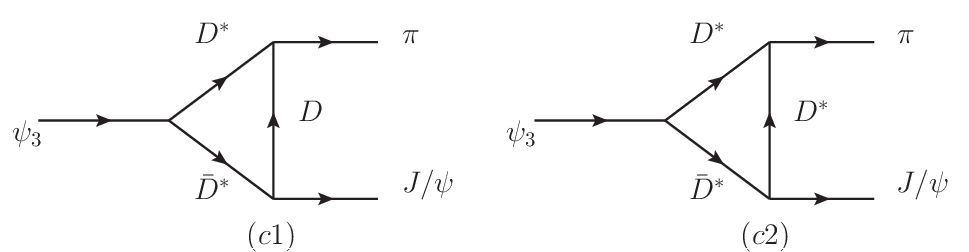}
\caption{Feynman diagrams contributing to $\psi(3842)\to J/\psi\pi$ at the hadron level.\label{figb2}}
\end{figure}

The meson loops contribute to $\psi(3842)\to J/\psi \pi$ are listed in Fig.~\ref{figb2}.
The amplitudes corresponding to Fig.~\ref{figb2} read,

\begin{equation}
\begin{array}{rl}
\mathcal{M}_{c1}= &i^3\displaystyle\int \dfrac{d^4 q}{(2\pi)^4}
 [g_{\psi_3 D^*D^*}(-1)\epsilon_{\psi_3}^{\mu\nu\alpha}(p)\\
&(-i)(g^{\nu\sigma}(p_1-p_2)^\mu g^{\alpha\delta} +g^{\alpha\sigma}(p_2-p_1)^\mu g^{\nu\delta})]\\
& [-i g_{D^*D\pi}(-1)(-i p_3)^\beta] \\
& [g_{\psi D^*D} \epsilon_{\kappa\tau\phi\xi} (-i p_4)^\kappa \epsilon_\psi^\tau(p_4) (-i)(-1)(q-p_2)^\phi ]\\
& \dfrac{-g^{\sigma\beta}+ p_1^\sigma p_1^\beta/m_{D^*}^2}{p_1^2-m_{D^*}^2}
\dfrac{-g^{\xi\delta}+ p_2^\xi p_2^\delta/m_{D^*}^2}{p_2^2-m_{D^*}^2}  \\
&\dfrac{1}{q^2-m_D^2}
\end{array}
\end{equation}

\begin{equation}
\begin{array}{rl}
\mathcal{M}_{c2}= &i^3\displaystyle\int \dfrac{d^4 q}{(2\pi)^4}
 [g_{\psi_3 D^*D^*}(-1)\epsilon_{\psi_3}^{\mu\nu\alpha}(p)\\
&(-i)(g^{\nu\sigma}(p_1-p_2)^\mu g^{\alpha\beta} +g^{\alpha\sigma}(p_2-p_1)^\mu g^{\nu\beta})]\\
&[\dfrac{g_{D^*D^*\pi} }{2} \epsilon_{\theta\tau\phi\rho}(-1)(-i p_3)^\tau(-i)(p_1+q)^\phi ] \\
& [i g_{\psi D^*D^*}(-i)(-1)(-\epsilon_{\psi_1}(p_4)\cdot (q-p_2)g^{\delta\chi} \\
&+\epsilon_\psi^\chi(p_4)(q-p_2)^\delta +\epsilon_\psi^\delta(p_4)(q-p_2)^\chi)]\\
&\dfrac{-g^{\rho\sigma}+ p_1^\sigma p_1^\rho/m_{D^*}^2}{p_1^2-m_{D^*}^2}
\dfrac{-g^{\beta\delta}+ p_2^\beta p_2^\delta/m_{D^*}^2}{p_2^2-m_{D^*}^2} \\
&\dfrac{-g^{\theta\chi}+ q^\theta q^\chi/m_{D^*}^2}{q^2-m_{D^*}^2}
\end{array}
\end{equation}

Then the total amplitude for $\psi(3842)\to J/\psi \pi$ read,
\begin{equation}
\label{eq1}
\begin{array}{rl}
\mathcal{M}^{tot}_{\psi(3842)\to h_c+\pi}= \displaystyle\sum^{2}_{i=1} 2\mathcal{M}_{ci} \mathcal{F}^2(m_q^2,q^2)
\end{array}
\end{equation}
with the factor $2$ in front of $\mathcal{M}_{ci}$ refers to the charge conjugate contribution. The amplitudes contributing to the $\psi(3842)\to J/\psi\pi$ process through $\eta-\pi$ mixing can be obtained through Eq.~(\ref{mix}).

\begin{figure}[htbp]
\includegraphics[width=8.2cm]{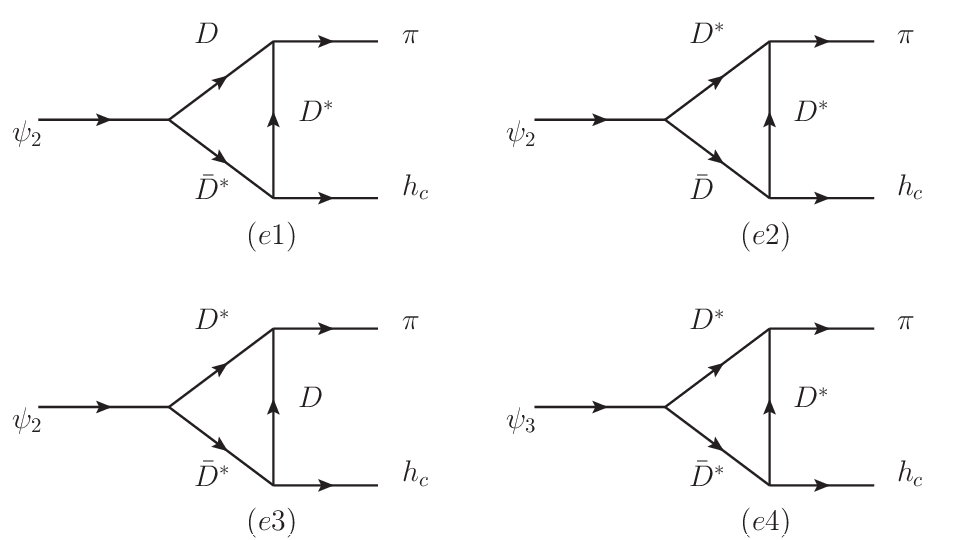}
\caption{Feynman diagrams contributing to $\psi(3823)\to h_c \pi$ at the hadron level.\label{figb3}}
\end{figure}

The meson loops contribute to $\psi(3823)\to h_c \pi$ are listed in Fig.~\ref{figb3}.
The amplitudes corresponding to Fig.~\ref{figb3} read,

\begin{equation}
\begin{array}{rl}
\mathcal{M}_{e1}= &i^3\displaystyle\int \dfrac{d^4 q}{(2\pi)^4}
[i g_{\psi_2 D^*D}   \epsilon_{\psi_2}^{\mu\nu}(p)(-1)(-i)(p_1-p_2)^\mu]\\
& [-i g_{D^*D\pi}(-i p_3)^\chi] [ g_{h_c D^* D^*}\epsilon_{\rho\sigma\alpha\beta}(-ip_4)^\rho \epsilon_{h_c}^\sigma(p_4) ]\\
&\dfrac{1}{p_1^2-m_D^2} \dfrac{-g^{\nu\beta}+ p_2^\nu p_2^\beta/m_{D^*}^2}{p_2^2-m_{D^*}^2}  \\
&\dfrac{-g^{\chi\alpha}+ q^\chi q^\alpha/m_{D^*}^2}{q^2-m_{D^*}^2}
\end{array}
\end{equation}

\begin{equation}
\begin{array}{rl}
\mathcal{M}_{e2}= &i^3\displaystyle\int \dfrac{d^4 q}{(2\pi)^4}
[i g_{\psi_2 D^*D} \epsilon_{\psi_2}^{\mu\nu}(p) (-i)(p_1-p_2)^\mu]\\
&[\dfrac{g_{D^*D^*\pi} }{2} \epsilon_{\chi\lambda\sigma\rho}(-1)(-i p_3)^\lambda(-i)(p_1+q)^\sigma ]\\
&[i g_{h_c D^*D} \epsilon_{h_c}^\alpha(p_4) ]\\
&\dfrac{-g^{\nu\rho}+ p_1^\nu p_1^\rho/m_{D^*}^2}{p_1^2-m_{D^*}^2} \dfrac{1}{p_2^2-m_D^2} \\
&\dfrac{-g^{\chi\alpha}+ q^\chi q^\alpha/m_{D^*}^2}{q^2-m_{D^*}^2}
\end{array}
\end{equation}

\begin{equation}
\begin{array}{rl}
\mathcal{M}_{e3}= &i^3\displaystyle\int \dfrac{d^4 q}{(2\pi)^4}
[g_{\psi_2 D^*D^*}\epsilon_{\lambda\nu\alpha\beta}(-1)(-i p^\lambda)\epsilon_{\psi_2}^{\mu\nu}(p)\\
&(-i)(g^{\alpha\sigma}(p_1-p_2)^\beta g^{\mu\delta} +g^{\mu\sigma}(p_2-p_1)^\beta g^{\alpha\delta})]\\
& [-i g_{D^*D\pi}(-1)(-i p_3)^\tau] [i g_{h_c D^*D} \epsilon_{h_c}^\kappa(p_4) ]\\
&\dfrac{-g^{\sigma\tau}+ p_1^\sigma p_1^\tau/m_{D^*}^2}{p_1^2-m_{D^*}^2}
\dfrac{-g^{\delta\kappa}+ p_2^\delta p_2^\kappa/m_{D^*}^2}{p_2^2-m_{D^*}^2} \\
&\dfrac{1}{q^2-m_D^2}
 \end{array}
\end{equation}

\begin{equation}
\begin{array}{rl}
\mathcal{M}_{e4}= &i^3\displaystyle\int \dfrac{d^4 q}{(2\pi)^4}
[g_{\psi_2 D^*D^*}\epsilon_{\lambda\nu\alpha\beta}(-1)(-i p^\lambda)\epsilon_{\psi_2}^{\mu\nu}(p)\\
&(-i)(g^{\alpha\sigma}(p_1-p_2)^\beta g^{\mu\delta} +g^{\mu\sigma}(p_2-p_1)^\beta g^{\alpha\delta})]\\
&[\dfrac{g_{D^*D^*\pi} }{2} \epsilon_{\chi\tau\phi\rho}(-1)(-i p_3)^\tau(-i)(p_1+q)^\phi g^{\rho\sigma}] \\
&[ g_{h_c D^* D^*}\epsilon_{\kappa\tau\xi\theta}(-ip_4)^\kappa \epsilon_{h_c}^\tau(p_4) ]\\
&\dfrac{-g^{\sigma\rho}+ p_1^\sigma p_1^\rho/m_{D^*}^2}{p_1^2-m_{D^*}^2}
\dfrac{-g^{\delta\theta}+ p_2^\delta p_2^\theta/m_{D^*}^2}{p_2^2-m_{D^*}^2} \\
&\dfrac{-g^{\chi\xi}+ q^\chi q^\xi/m_{D^*}^2}{q^2-m_{D^*}^2}
\end{array}
\end{equation}

Then the total amplitude for $\psi(3823)\to h_c \pi$ read,
\begin{equation}
\label{eq1}
\begin{array}{rl}
\mathcal{M}^{tot}_{\psi(3823)\to h_c\pi}= \displaystyle\sum^{4}_{i=1} 2\mathcal{M}_{ei} \mathcal{F}^2(m_q^2,q^2)
\end{array}
\end{equation}
with the factor $2$ in front of $\mathcal{M}_{ei}$ refers to the charge conjugate contribution. The amplitudes contributing to the $\psi(3823)\to h_c\pi$ process through $\eta-\pi$ mixing can be obtained through Eq.~(\ref{mix}).

\begin{figure}[htbp]
\includegraphics[width=8.2cm]{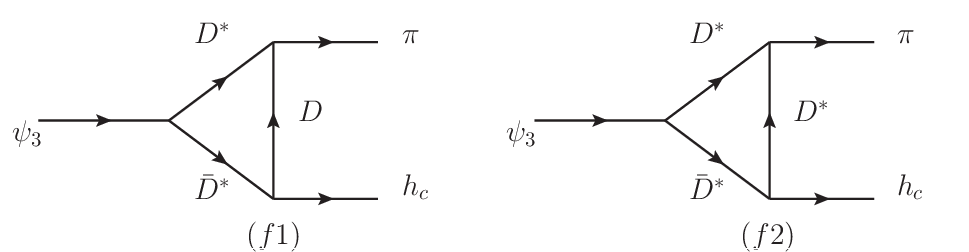}
\caption{Feynman diagrams contributing to $\psi(3842)\to h_c \pi$ at the hadron level.\label{figb4}}
\end{figure}

The meson loops contribute to $\psi(3842)\to h_c \pi$ are listed in Fig.~\ref{figb4}.
The amplitudes corresponding to Fig.~\ref{figb4} read,

\begin{equation}
\begin{array}{rl}
\mathcal{M}_{f1}= &i^3\displaystyle\int \dfrac{d^4 q}{(2\pi)^4} [g_{\psi_3 D^*D^*}(-1)\epsilon_{\psi_3}^{\mu\nu\alpha}(p)\\
&(-i)(g^{\nu\sigma}(p_1-p_2)^\mu g^{\alpha\delta} +g^{\alpha\sigma}(p_2-p_1)^\mu g^{\nu\delta})]\\
& [-i g_{D^*D\pi}(-1)(-i p_3)^\rho] [i g_{h_c D^*D} \epsilon_{h_c}^\tau(p_4) ]\\
&\dfrac{-g^{\sigma\rho}+ p_1^\sigma p_1^\rho/m_{D^*}^2}{p_1^2-m_{D^*}^2}
\dfrac{-g^{\delta\tau}+ p_2^\delta p_2^\tau/m_{D^*}^2}{p_2^2-m_{D^*}^2} \\
&\dfrac{1}{q^2-m_D^2}
\end{array}
\end{equation}

\begin{equation}
\begin{array}{rl}
\mathcal{M}_{f2}= &i^3\displaystyle\int \dfrac{d^4 q}{(2\pi)^4} [g_{\psi_3 D^*D^*}(-1)\epsilon_{\psi_3}^{\mu\nu\alpha}(p)\\
&(-i)(g^{\nu\sigma}(p_1-p_2)^\mu g^{\alpha\delta} +g^{\alpha\sigma}(p_2-p_1)^\mu g^{\nu\delta})]\\
&[\dfrac{g_{D^*D^*\pi} }{2} \epsilon_{\chi\tau\phi\rho}(-1)(-i p_3)^\tau(-i)(p_1+q)^\phi ] \\
&[ g_{h_c D^* D^*}\epsilon_{\kappa\lambda\xi\theta}(-ip_4)^\kappa \epsilon_{h_c}^\lambda(p_4) ]\\
&\dfrac{-g^{\sigma\rho}+ p_1^\sigma p_1^\rho/m_{D^*}^2}{p_1^2-m_{D^*}^2}
\dfrac{-g^{\delta\theta}+ p_2^\delta p_2^\theta/m_{D^*}^2}{p_2^2-m_{D^*}^2}\\
& \dfrac{-g^{\chi\xi}+ q^\chi q^\xi/m_{D^*}^2}{q^2-m_{D^*}^2}
\end{array}
\end{equation}

Then the total amplitude for $\psi(3842)\to h_c \pi$ read,
\begin{equation}
\label{eq1}
\begin{array}{rl}
\mathcal{M}^{tot}_{\psi(3842)\to h_c\pi}= \displaystyle\sum^{2}_{i=1} 2\mathcal{M}_{fi} \mathcal{F}^2(m_q^2,q^2)
\end{array}
\end{equation}
with the factor $2$ in front of $\mathcal{M}_{fi}$ refers to the charge conjugate contribution. The amplitudes contributing to the $\psi(3842)\to h_c\pi$ process through $\eta-\pi$ mixing can be obtained through Eq.~(\ref{mix}).


\clearpage


\begin{thebibliography}{99}

\bibitem{Rapidis:1977cv}
P.~A.~Rapidis, B.~Gobbi, D.~Luke, A.~Barbaro-Galtieri, J.~Dorfan, R.~Ely, G.~J.~Feldman, J.~M.~Feller, A.~Fong and G.~Hanson, \textit{et al.}
Phys. Rev. Lett. \textbf{39}, 526 (1977)
[erratum: Phys. Rev. Lett. \textbf{39}, 974 (1977)]

\bibitem{Rosner:2001nm}
J.~L.~Rosner,
Phys. Rev. D \textbf{64} (2001), 094002

\bibitem{Voloshin:2005sd}
M.~B.~Voloshin,
Phys. Rev. D \textbf{71} (2005), 114003

\bibitem{BESIII:2023oql}
M.~Ablikim \textit{et al.} [BESIII],
Phys. Rev. Lett. \textbf{133} (2024) no.24, 241902

\bibitem{Rosner:2004wy}
J.~L.~Rosner,
Annals Phys. \textbf{319} (2005), 1-12

\bibitem{Rosner:2004mi}
J.~L.~Rosner,
[arXiv:hep-ph/0405196 [hep-ph]].

\bibitem{ParticleDataGroup:2024cfk}
S.~Navas \textit{et al.} [Particle Data Group],
Phys. Rev. D \textbf{110}, no.3, 030001 (2024)

\bibitem{CLEO:2007rrw}
S.~Dobbs \textit{et al.} [CLEO],
Phys. Rev. D \textbf{76}, 112001 (2007)



\bibitem{CLEO:2005pql}
Q.~He \textit{et al.} [CLEO],
Phys. Rev. Lett. \textbf{95}, 121801 (2005)
[erratum: Phys. Rev. Lett. \textbf{96}, 199903 (2006)]



\bibitem{Kuang:1989ub}
Y.~P.~Kuang and T.~M.~Yan,
Phys. Rev. D \textbf{41} (1990), 155

\bibitem{Ding:1991vu}
Y.~B.~Ding, D.~H.~Qin and K.~T.~Chao,
Phys. Rev. D \textbf{44} (1991), 3562-3566

\bibitem{Eichten:2007qx}
E.~Eichten, S.~Godfrey, H.~Mahlke and J.~L.~Rosner,
Rev. Mod. Phys. \textbf{80} (2008), 1161-1193

\bibitem{BES:2003bes}
J.~Z.~Bai \textit{et al.} [BES],
Phys. Lett. B \textbf{605} (2005), 63-71

\bibitem{CLEO:2005zky}
N.~E.~Adam \textit{et al.} [CLEO],
Phys. Rev. Lett. \textbf{96} (2006), 082004

\bibitem{CLEO:2005vma}
T.~E.~Coan \textit{et al.} [CLEO],
Phys. Rev. Lett. \textbf{96} (2006), 182002

\bibitem{CLEO:2006nor}
R.~A.~Briere \textit{et al.} [CLEO],
Phys. Rev. D \textbf{74} (2006), 031106

\bibitem{CLEO:2005zrs}
G.~S.~Adams \textit{et al.} [CLEO],
Phys. Rev. D \textbf{73} (2006), 012002

\bibitem{BES:2006fpf}
M.~Ablikim \textit{et al.} [BES],
Phys. Rev. Lett. \textbf{97} (2006), 121801

\bibitem{BES:2006dso}
M.~Ablikim \textit{et al.} [BES],
Phys. Lett. B \textbf{641} (2006), 145-155

\bibitem{BES:2007cev}
M.~Ablikim \textit{et al.} [BES],
Phys. Rev. D \textbf{76} (2007), 122002

\bibitem{BES:2008vad}
M.~Ablikim \textit{et al.} [BES],
Phys. Lett. B \textbf{659} (2008), 74-79



\bibitem{He:2008xb}
Z.~G.~He, Y.~Fan and K.~T.~Chao,
Phys. Rev. Lett. \textbf{101} (2008), 112001

\bibitem{Achasov:2005qb}
N.~N.~Achasov and A.~A.~Kozhevnikov,
Phys. Atom. Nucl. \textbf{69} (2006), 988-998



\bibitem{CLEO:2005mpm}
D.~Besson \textit{et al.} [CLEO],
Phys. Rev. Lett. \textbf{96} (2006), 092002
[erratum: Phys. Rev. Lett. \textbf{104} (2010), 159901]

\bibitem{Achasov:1990gt}
N.~N.~Achasov and A.~A.~Kozhevnikov,
Phys. Lett. B \textbf{260} (1991), 425-428

\bibitem{Achasov:1994vh}
N.~N.~Achasov and A.~A.~Kozhevnikov,
Phys. Rev. D \textbf{49} (1994), 275-282

\bibitem{Ioffe:1980mx}
B.~L.~Ioffe and M.~A.~Shifman,
Phys. Lett. B \textbf{95} (1980), 99-102

\bibitem{Zhang:2009kr}
Y.~J.~Zhang, G.~Li and Q.~Zhao,
Phys. Rev. Lett. \textbf{102} (2009), 172001

\bibitem{Liu:2009dr}
X.~Liu, B.~Zhang and X.~Q.~Li,
Phys. Lett. B \textbf{675} (2009), 441-445

\bibitem{Guo:2012tj}
Z.~k.~Guo, S.~Narison, J.~M.~Richard and Q.~Zhao,
Phys. Rev. D \textbf{85}, 114007 (2012)

\bibitem{Li:2013zcr}
G.~Li, X.~h.~Liu, Q.~Wang and Q.~Zhao,
Phys. Rev. D \textbf{88}, no.1, 014010 (2013)

\bibitem{Wang:2011yh}
Q.~Wang, X.~H.~Liu and Q.~Zhao,
Phys. Rev. D \textbf{84}, 014007 (2011)


\bibitem{Bai:2026atm}
Z.~Y.~Bai, D.~Y.~Chen, Q.~Huang, X.~Liu, S.~Q.~Luo and J.~Z.~Wang,
Phys. Rept. \textbf{1204} (2027), 1-162

\bibitem{BESIII:2026pyf}
M.~Ablikim \textit{et al.} [BESIII],
[arXiv:2606.14105 [hep-ex]].








\bibitem{Belle:2013ewt}
V.~Bhardwaj \textit{et al.} [Belle],
Phys. Rev. Lett. \textbf{111}, no.3, 032001 (2013)

\bibitem{LHCb:2019lnr}
R.~Aaij \textit{et al.} [LHCb],
JHEP \textbf{07}, 035 (2019)

\bibitem{Qi:2025hwd}
X.~Y.~Qi, Q.~Wu, X.~D.~Guo and D.~Y.~Chen,
Eur. Phys. J. C \textbf{85}, no.1, 96 (2025)

\bibitem{Qi:2025xkz}
X.~Y.~Qi, X.~D.~Guo and D.~Y.~Chen,
Phys. Rev. D \textbf{112}, no.5, 054003 (2025)






\bibitem{Mannel:1997ky}
T.~Mannel,
Rept. Prog. Phys. \textbf{60}, 1113-1172 (1997)

\bibitem{Brambilla:2004jw}
N.~Brambilla, A.~Pineda, J.~Soto and A.~Vairo,
Rev. Mod. Phys. \textbf{77}, 1423 (2005)

\bibitem{Colangelo:2003sa}
P.~Colangelo, F.~De Fazio and T.~N.~Pham,
Phys. Rev. D \textbf{69}, 054023 (2004)

\bibitem{Lipkin:1986av}
H.~J.~Lipkin,
Phys. Lett. B \textbf{179}, 278 (1986)

\bibitem{Casalbuoni:1996pg}
R.~Casalbuoni, A.~Deandrea, N.~Di Bartolomeo, R.~Gatto, F.~Feruglio and G.~Nardulli,
Phys. Rept. \textbf{281}, 145-238 (1997)

\bibitem{Kaymakcalan:1983qq}
O.~Kaymakcalan, S.~Rajeev and J.~Schechter,
Phys. Rev. D \textbf{30}, 594 (1984)

\bibitem{Oh:2000qr}
Y.~s.~Oh, T.~Song and S.~H.~Lee,
Phys. Rev. C \textbf{63}, 034901 (2001)

\bibitem{Falk:1992cx}
A.~F.~Falk and M.~E.~Luke,
Phys. Lett. B \textbf{292}, 119-127 (1992)

\bibitem{Chen:2014sra}
D.~Y.~Chen, X.~Liu and T.~Matsuki,
Phys. Rev. D \textbf{91}, no.9, 094023 (2015)

\bibitem{Yan:1992gz}
T.~M.~Yan, H.~Y.~Cheng, C.~Y.~Cheung, G.~L.~Lin, Y.~C.~Lin and H.~L.~Yu,
Phys. Rev. D \textbf{46}, 1148-1164 (1992)
[erratum: Phys. Rev. D \textbf{55}, 5851 (1997)]

\bibitem{Cheng:1992xi}
H.~Y.~Cheng, C.~Y.~Cheung, G.~L.~Lin, Y.~C.~Lin, T.~M.~Yan and H.~L.~Yu,
Phys. Rev. D \textbf{47}, 1030-1042 (1993)

\bibitem{Wise:1992hn}
M.~B.~Wise,
Phys. Rev. D \textbf{45}, no.7, R2188 (1992)

\bibitem{Casalbuoni:1992gi}
R.~Casalbuoni, A.~Deandrea, N.~Di Bartolomeo, R.~Gatto, F.~Feruglio and G.~Nardulli,
Phys. Lett. B \textbf{292}, 371-376 (1992)

\bibitem{Casalbuoni:1992dx}
R.~Casalbuoni, A.~Deandrea, N.~Di Bartolomeo, R.~Gatto, F.~Feruglio and G.~Nardulli,
Phys. Lett. B \textbf{299}, 139-150 (1993)

\bibitem{Gilman:1987ax}
F.~J.~Gilman and R.~Kauffman,
Phys. Rev. D \textbf{36}, 2761 (1987)
[erratum: Phys. Rev. D \textbf{37}, 3348 (1988)]

\bibitem{MARK-III:1988crp}
D.~Coffman \textit{et al.} [MARK-III],
Phys. Rev. D \textbf{38}, 2695 (1988)
[erratum: Phys. Rev. D \textbf{40}, 3788 (1989)]

\bibitem{DM2:1988bfq}
J.~Jousset \textit{et al.} [DM2],
Phys. Rev. D \textbf{41}, 1389 (1990)

\bibitem{Xiao:2016hoa}
C.~J.~Xiao, D.~Y.~Chen and Y.~L.~Ma,
Phys. Rev. D \textbf{93}, no.9, 094011 (2016)

\bibitem{Yue:2025wcl}
Z.~L.~Yue, Q.~Y.~Guo, D.~Y.~Chen and E.~Santopinto,
Eur. Phys. J. C \textbf{86}, no.1, 33 (2026)


\bibitem{Tornqvist:1993vu}
N.~A.~Tornqvist,
Nuovo Cim. A \textbf{107}, 2471-2476 (1994)

\bibitem{Tornqvist:1993ng}
N.~A.~Tornqvist,
Z. Phys. C \textbf{61}, 525-537 (1994)

\bibitem{Locher:1993cc}
M.~P.~Locher, Y.~Lu and B.~S.~Zou,
Z. Phys. A \textbf{347}, 281-284 (1994)

\bibitem{Li:1996yn}
X.~Q.~Li, D.~V.~Bugg and B.~S.~Zou,
Phys. Rev. D \textbf{55}, 1421-1424 (1997)

\bibitem{Zhang:2009gy}
Y.~J.~Zhang and Q.~Zhao,
Phys. Rev. D \textbf{81}, 034011 (2010)

\bibitem{Qi:2023kwc}
X.~Y.~Qi, Q.~Wu and D.~Y.~Chen,
Eur. Phys. J. C \textbf{83}, no.11, 1006 (2023)

\bibitem{Wu:2025crk}
Q.~Wu, Z.~Q.~Sun, D.~Y.~Chen, S.~D.~Liu and G.~Li,
Phys. Rev. D \textbf{113}, no.1, 014015 (2026)







\end{thebibliography}
\end{document}